\documentclass[fleqn,usenatbib]{mnras}

\usepackage{mhchem}
\usepackage{newtxtext,newtxmath}
\usepackage{xcolor}
\usepackage{xspace}
\usepackage{multicol}
\usepackage{multirow}

\usepackage[T1]{fontenc}

\DeclareRobustCommand{\VAN}[3]{#2}
\let\VANthebibliography\thebibliography
\def\thebibliography{\DeclareRobustCommand{\VAN}[3]{##3}\VANthebibliography}

\usepackage{graphicx}	
\usepackage{amsmath}	
\usepackage{booktabs}

\def\tablefoot#1{\par\vspace*{2ex}%
 \parbox{\hsize}{\leftskip0pt\rightskip0pt
 {\noindent\small\textbf{Notes.}~#1\par}}}

\newcommand{\Tabref}[1]{Table~\ref{#1}}

\renewcommand{\eqref}[1]{Eq.~(\ref{#1})}
\newcommand{\Eqref}[1]{Equation~(\ref{#1})}

\newcommand{\Reacref}[1]{R.\ref{#1}}

\newcommand{\Secref}[1]{Section~\ref{#1}}

\newcommand{\one}{\textbf{1}\xspace}
\newcommand{\two}{\textbf{2}\xspace}
\newcommand{\kcalmol}{kcal mol$^{-1}$\xspace}
\title[HOCO Chemistry]{Hydrogenation of HOCO and formation of interstellar \ce{CO2}: A not so straightforward relation}

\author[G. Molpeceres et al.]{
Germ\'an Molpeceres,$^{1}$\thanks{E-mail: german.molpeceres@iff.csic.es}
Joan Enrique-Romero,$^{2}$\thanks{E-mail: j.enrique.romero@umail.leidenuniv.nl},\thanks{Both corresponding authors contributed equally to this work},
Atsuki Ishibashi,$^{3}$
Yasuhiro Oba,$^{4}$
Hiroshi Hidaka,$^{4}$
\newauthor
Thanja Lamberts,$^{2,5}$
Yuri Aikawa,$^{6}$
Naoki Watanabe.$^{4}$
\\
$^{1}$Departamento de Astrofísica Molecular, Instituto de F\'isica Fundamental, CSIC, C/ Serrano 123,113bis,121, 28006 Madrid, Spain\\
$^{2}$Leiden Institute of Chemistry, Leiden University, Leiden 2300 RA, The Netherlands\\
$^{3}$Komaba Institute for Science, The University of Tokyo, 4-6-1 Komaba, Meguro, Tokyo 153-8902, Japan\\
$^{4}$Institute of Low Temperature Science,Hokkaido University, Sapporo, Hokkaido 060-0819, Japan \\
$^{5}$Leiden Observatory, Leiden University, P.O. Box 9513, 2300 RA Leiden, The Netherlands\\
$^{6}$Department of Astronomy, Graduate School of Science, The University of Tokyo, Tokyo 113 0033, Japan
}

\date{Accepted XXX. Received YYY; in original form ZZZ}

\pubyear{2024}

\begin{document}
\label{firstpage}
\pagerange{\pageref{firstpage}--\pageref{lastpage}}
\maketitle

\begin{abstract}
Carbon dioxide (\ce{CO2}) is one of the most important interstellar molecules. While it is considered that it forms on the surface of interstellar dust grains, the exact contribution of different chemical mechanisms is still poorly constrained. Traditionally it is deemed that the \ce{CO + OH} reaction occurring on top of ices is the main reaction path for its formation. Recent investigations showed that in reality the reaction presents a more complex mechanism, requiring an additional H-abstraction step. Building on our previous works, we carried out a detailed investigation of such H abstraction reactions with the hydrogen atom as a reactant for the abstraction reaction. We found an unconventional chemistry for this reaction, markedly depending on the isomeric form of the HOCO radical prior to reaction. The favored reactions are \ce{t-HOCO + H -> CO + H2O}, \ce{c-HOCO + H -> CO2 + H2} and \ce{t/c-HOCO + H -> c/t-HCOOH}. We estimate bounds for the rate constants of the less favored reaction channels, \ce{t-HOCO + H -> CO2 + H} and \ce{c-HOCO + H -> CO + H2O}, to be approximately 10$^{4-6}$ s$^{-1}$. However, these estimates should be interpreted cautiously due to the significant role of quantum tunneling in these reactions and the complex electronic structure of the involved molecules, which complicates their study. Our findings underscore the need for detailed investigation into the chemistry of interstellar \ce{CO2} and pave the way for a reevaluation of its primary formation mechanisms in the interstellar medium. 
\end{abstract}

\begin{keywords}
    ISM: molecules -- molecular data -- astrochemistry -- methods: numerical
\end{keywords}



\section{Introduction}

\allowdisplaybreaks

Carbon dioxide (\ce{CO2}) is one of the most important molecules in astrophysics and beyond. One of its formation routes,

\begin{equation}
    \ce{CO + OH -> CO2 + H} \label{eq:co2},
\end{equation}

\noindent has sometimes been labelled as the ``second most important'' reaction in combustion chemistry \citep{masunov_catalytic_2018}. From an astrochemical perspective the low temperature behavior of the reaction is more interesting, at least in interstellar dark clouds. In the gas phase, the reaction is reported to be slow, with experimentally determined rate constants on the order of 10$^{-13}$ cm$^{3}$ s$^{-1}$ at 300 K \citep{Frost1991, greenblatt_oxygen_1989} and even lower from theoretical extrapolations to lower temperatures \citep{lakin_quasiclassical_2003, senosiain_complete_2005,Ma2012,caracciolo_combined_2018,li_quantum_2014}. This, in turn, shifted the focus towards the formation of interstellar \ce{CO2} to experiments where the reaction happens on the surface of ice coated dust grains \citep{ioppolo_surface_2011,noble_co_2011,Oba2010Carbonic,oba_experimental_2010,Qasim2019,terwisscha_van_scheltinga_formation_2022}. All these experiments demonstrated the formation of \ce{CO2}, reinforcing the feasibility of \Reacref{eq:co2} on interstellar ices and prompting models to adopt minimal activation energies for \Reacref{eq:co2} \citep[e.g.,][]{Garrod2011,Pauly2018,clement_astrochemical_2023}. 

Recently, we conducted a theoretical study on \Reacref{eq:co2} considering two different substrates, \ce{H2O} and \ce{CO}, and different energy dissipation scenarios \citep{molpeceres_cracking_2023}. The goal of our study was twofold: first, to determine whether the ice matrix exerts a catalytic effect on the formation of \ce{CO2} ice; and second, to investigate the role of energy dissipation in facilitating the reaction. We concluded that \Reacref{eq:co2} was very inefficient on ice surfaces. There was no evidence of catalysis through the ice matrix and, even in the best case scenario, with minimal energy dissipation, the reaction was very slow. Our theoretical claims were confirmed by new, highly sensitive, experimental measurements \citep{ishibashi_proposed_2024}. Therefore, a valid question arises: Why there seems to be a consensus that \Reacref{eq:co2} produces \ce{CO2} in ice experiments? The answer to that question lies in the mechanism for the reaction. Unlike most other reactions on interstellar ices, \Reacref{eq:co2} is not elemental, meaning that several reaction steps in a reaction mechanism are needed. Notably, the reaction stops after the formation of HOCO, a reactive radical that presents cis-trans conformerism (c-HOCO/t-HOCO). In all previous experimental works on ices, the ice contained a significant amount of other radicals, e.g. OH and H that could lead to \ce{CO2} via a subsequent H-atom abstraction reaction. Because the experimental techniques used in those studies were primarily suited for identifying final products, and do not necessarily simulate the actual interstellar environment, it was challenging to disentangle the dominant reaction pathway or mechanism. These limitations are lifted in the theoretical research for obvious reasons, but also in our newest experiments \citep{ishibashi_proposed_2024}. 

After determining that \Reacref{eq:co2} is not an elementary reaction and halts at (c/t)-HOCO, we proposed the following abstraction reaction:

\begin{equation}
    \ce{HOCO + H -> CO2 + H2}, \label{eq:naive_abstraction}
\end{equation}

\noindent as the subsequent reaction for the formation of \ce{CO2}. However, it is clear that there is more than one reaction channel for the reaction of HOCO \citep{francisco_hoco_2010}, which is especially true on ice surfaces, where the addition product (formic acid, \ce{HCOOH}) can thermalize. Because HOCO presents cis-trans isomerism (see Figure \ref{fig:cis_and_tras}), there are twice as many reaction channels. Therefore, in reality, \Reacref{eq:naive_abstraction} can be decomposed as:

\begin{figure}
    \centering
	\includegraphics[width=0.15\textwidth]{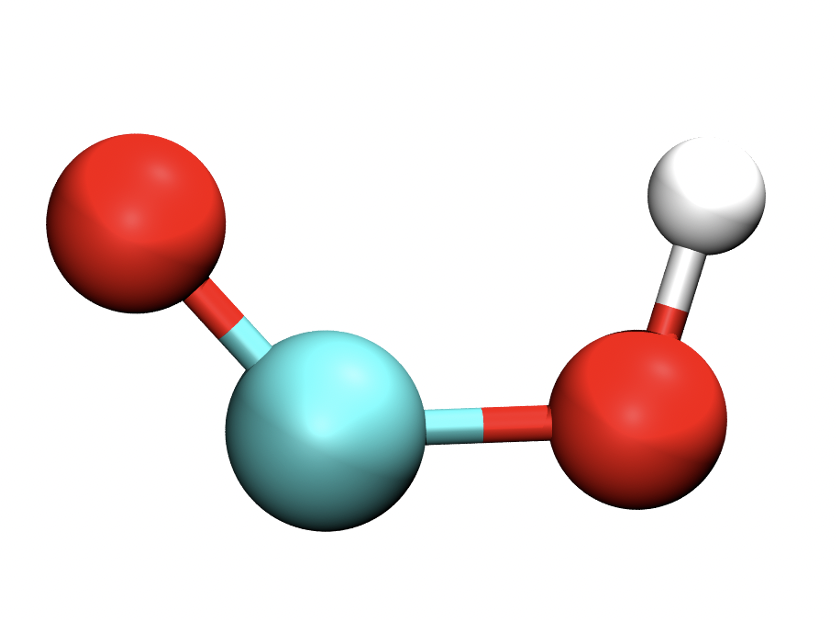}\\
	\textbf{(a)} cis-HOCO\\
	\includegraphics[width=0.15\textwidth]{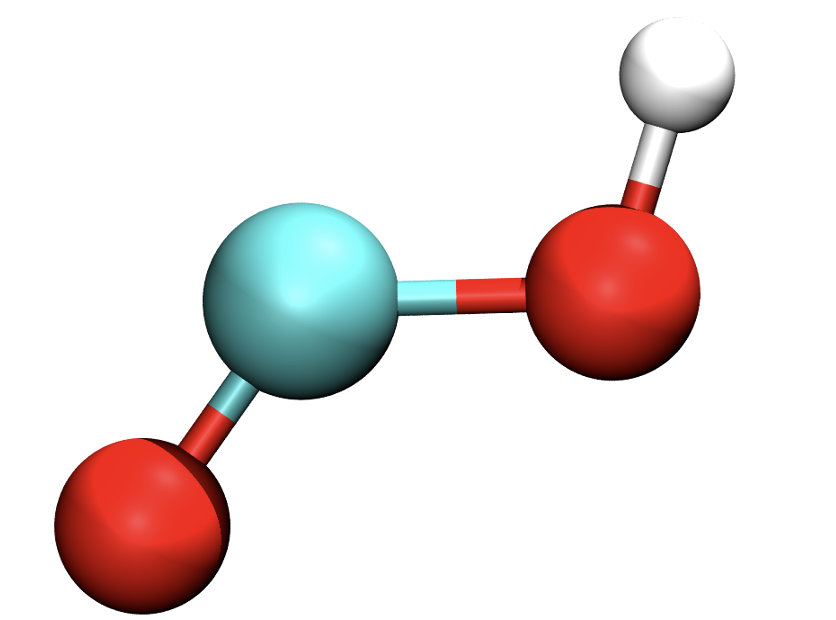}\\
	\textbf{(b)} trans-HOCO
	\caption{The HOCO radical in its cis and trans conformations. Color code for this figure and ones below is red for oxygen, gray for carbon and white for hydrogen.}
	\label{fig:cis_and_tras}
\end{figure}

\begin{align}
    \ce{t-HOCO + H &-> CO2 + H2} \label{eq:t_co2} \\
    \ce{c-HOCO + H &-> CO2 + H2} \label{eq:c_co2} \\
    \ce{t-HOCO + H &-> CO + H2O} \label{eq:t_co}  \\
    \ce{c-HOCO + H &-> CO + H2O} \label{eq:c_co}  \\
    \ce{t-HOCO + H &-> c-HCOOH} \label{eq:t2_hcooh} \\
    \ce{c-HOCO + H &-> t-HCOOH} \label{eq:c2_hcooh}.
\end{align}

\noindent Furthermore, because the cis-trans isomerism of HOCO determines the outcome of reactions \Reacref{eq:t2_hcooh} and \Reacref{eq:c2_hcooh}, H-abstraction in reactions in HCOOH are also an intrinsic part of the reaction network with different products:

\begin{align}
    \ce{t-HCOOH + H &-> c-HOCO + H2} \label{eq:h-abs_t} \\
    \ce{c-HCOOH + H &-> t-HOCO + H2} \label{eq:h-abs_c}.
\end{align}

\noindent Reactions \Reacref{eq:h-abs_t} and \Reacref{eq:h-abs_c} were studied recently \citep{Molpeceres2022Thio} by some of us finding a major influence of HCOOH isomerism in the reaction. Therefore, it is clear that \Reacref{eq:naive_abstraction} is not trivial, despite being a radical-radical reaction, where, in principle, kinetic barriers should be lower than in reactions involving a radical and a closed-shell molecule.

In this work, we aim to computationally explore the details of \Reacref{eq:naive_abstraction}, extending our investigation to examine the role of competing reaction channels and the HOCO isomer prior to the reaction. To our surprise, we discovered that these two factors have a major impact on the reaction outcome in unexpected ways, highlighting the need to reconsider the chain reactions \Reacref{eq:co2} + \Reacref{eq:naive_abstraction} as the primary mechanism for producing interstellar \ce{CO2} at cryogenic temperatures (10 K).

This article is structured as follows. In \Secref{sec:methods} we describe the computational setting for the study of the HOCO reactivity, including an ample benchmark of electronic structure methods. In \Secref{sec:results} we show our computational results, including potential energy surface (PES) investigations for each reaction and a tentative kinetic analysis. \Secref{sec:discussion} delves into the chemical rationale of our findings, the limitations of our study and needed subsequent works. Finally, \Secref{sec:conclusions} provides a brief summary of our results along with our main conclusions.

\section{Methodology} \label{sec:methods}

\subsection{Benchmark} \label{sec:method:bs}

The computational chemistry codes employed in our electronic structure calculations include {\sc Orca} (v.6.0.0) \cite{Neese2020,Neese2022} for density functional theory (DFT) calculations, and the {\sc OpenMolcas} suite \citep{fdez_galvan_openmolcas_2019, aquilante_modern_2020} for multireference calculations.

The study of the hydrogenation of the HOCO radical has two requirements. In the first place, the \ce{HOCO + H} reaction involves the recombination of two radicals, each in a doublet state. From a computational chemistry point of view, the recombination in the global singlet channel, i.e. two antiparallel spins, is the most interesting one for reactivity. Such a recombination channel is intrinsically multireference, where with wave function of the system described by a combination of electronic configurations. More details on this issue in radical chemistry can be found in \citet{enrique-romero_revisiting_2020}. Here it suffices to say that, while in many cases a qualitative picture is sufficient \citep{molpeceres_carbon_2024}, in others where accurate activation energies $\Delta U_{A}$ are needed, density functional theory (DFT) can only be safely used after proper benchmarking. To accurately benchmark which functionals perform best for the reactions under study, we have employed CASPT2 \citep{andersson_second-order_1990} and XMS-CASPT2 (\cite{granovsky_extended_2011}; expanding the first five electronic states) single point calculations as references. This has severely limited the number of atoms used in the benchmark study, since we could not include more than the reacting atoms (HOCO + H).  To obtain the initial geometries, we cut out the transition state structures for reactions \Reacref{eq:t_co2}--\Reacref{eq:c_co} on a water ice model, producing geometries where only the reacting molecules are present. With these extracted geometries we explored the potential energy surface (PES) for each reaction in the ``gas-phase'' (i.e., without any water surface) with MPWB1K-D3(BJ)/def2-TZVPPD \citep{Zhao2005,grimme2010, Grimme2011}. We could only find transition states (TS) for reactions \Reacref{eq:t_co2} and \Reacref{eq:c_co}, since the others were found to be barrierless. The selection of MPWB1K-D3(BJ)/def2-TZVPPD as the initial exploratory method is based on the excellent behavior found for the reactivity of a related molecule, formic acid (\cite{Molpeceres2021carbon}), although the method was not ultimately selected as the best one for the chemistry of the HOCO radical. In addition, CASPT2 was found to be a poor reference for reaction \Reacref{eq:t_co2}, since the two first electronic states of this system appear to be very close at the TS geometry, about 2.4 \kcalmol apart, while for the other TSs it was above 71 \kcalmol, hence the XMS-CASPT2 method was used instead. The basis set for the reference calculations was set to cc-pVTZ \citep{Duning1989}, and for the functionals being tested it was set to be def2-TZVPPD \citep{rappoport_property-optimized_2010}. The active space employed in the CASPT and XMS-CASPT2 calculations involves 18 active electrons and 14 molecular orbitals, which contain all the valence electrons and orbitals. All the DFT calculations were carried out using a broken symmetry formalism to ensure the convergence to a biradical wavefunction.

\begin{figure*}
	\centering
	\includegraphics[width=0.8\textwidth]{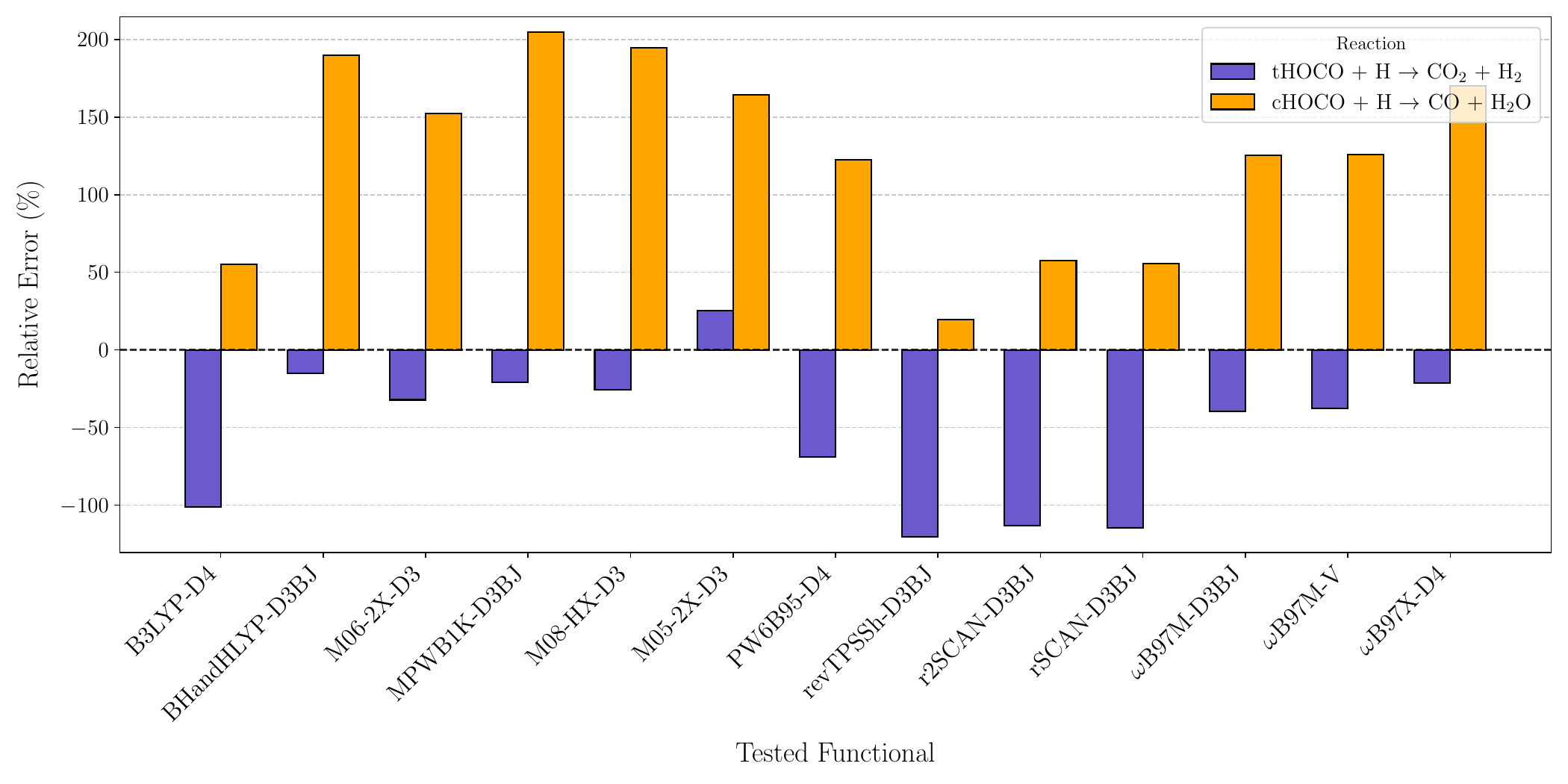}
	\caption{Relative signed error (\%) between singlet point energies for different functionals and the activation energy barriers for reactions 
    \ce{tHOCO + H -> CO2 + H2} (\Reacref{eq:t_co2}; 12.2 kcal mol$^{-1}$) and \ce{cHOCO + H -> CO + H2O} (\Reacref{eq:c_co}; 5.3 kcal mol$^{-1}$). Reactions \ce{tHOCO + H -> CO + H2O} (\Reacref{eq:t_co}) and \ce{cHOCO + H -> CO2 + H2} (\Reacref{eq:c_co2}) where found to be barrierless. Find further details on the reasons for these finding the main text (Section \ref{sec:ration}).}
	\label{fig:benchmark}
\end{figure*}

The benchmark results are presented in Figure \ref{fig:benchmark}, which includes testing of 13 exchange-correlation functionals. For reactions \Reacref{eq:c_co2}--\Reacref{eq:c_co}, revTPSSh-D3(BJ) \citep{perdew_workhorse_2009, grimme2010, Grimme2011} emerged as one of the top performers. In contrast, for reaction \ref{eq:t_co2}, the BHandHLYP-D3(BJ) \citep{Becke1993, grimme2010, Grimme2011} functional was selected for its simplicity and its close agreement with the energy reference. In the following, and unless stated otherwise, level \one corresponds to the BHandHLYP(D3BJ)/def2-TZVPPD//BHandHLYP(D3BJ)-gCP/def2-SVP and level \two to revTPSSh(D3BJ)/def2-TZVPPD//revTPSSh(D3BJ)-gCP/def2-SVP. The gCP suffix to the method indicates the the geometric counterpoise correction \citep{Sure2013} was applied to the geometry optimization and Hessian calculations to palliate the basis set superposition error stemming from the reduced basis set for the calculations with the ice model.  

\subsection{Reactions on an ice cluster} \label{sec:method:cluster}

\begin{figure}
	\centering
	\includegraphics[width=0.4\columnwidth]{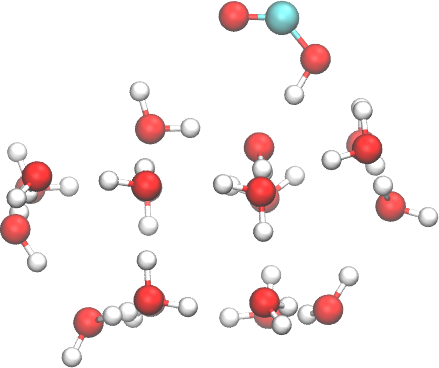}
    \hfill
    \includegraphics[width=0.4\columnwidth]{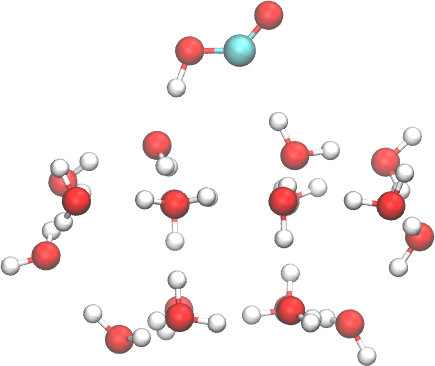}\\
    \textbf{(a)} Site I

    \vfill
    \includegraphics[width=0.4\columnwidth]{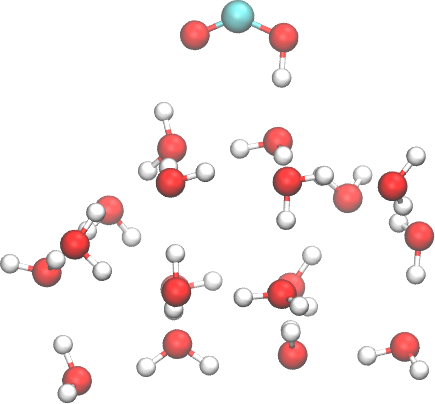}
    \hfill
    \includegraphics[width=0.4\columnwidth]{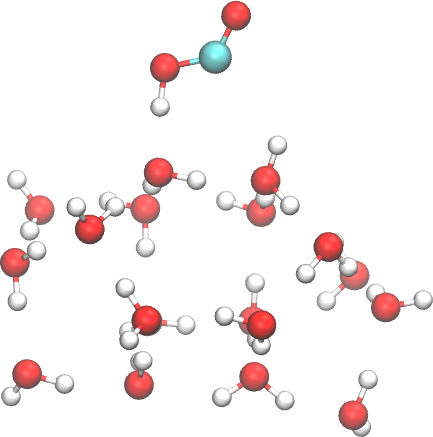}\\
    \textbf{(b)} Site II
    \vfill

    \includegraphics[width=0.4\columnwidth]{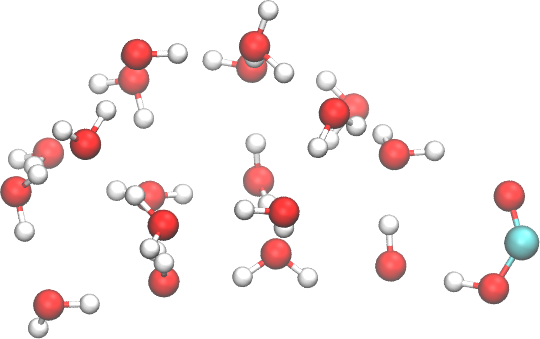}
    \hfill
    \includegraphics[width=0.4\columnwidth]{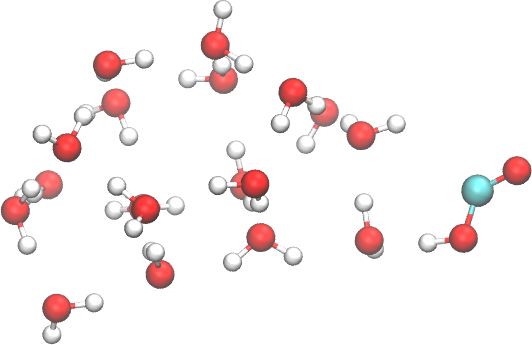}\\
    \textbf{(c)} Site III

	\caption{Depiction of the binding sites for c-HOCO and t-HOCO from which we start our simulations. All geometries correspond to optimizations carried out with the revTPSSh(D3BJ)-gCP/def2-SVP method. Left column corresponds to c-HOCO and right column to t-HOCO. }
	\label{fig:sites}
\end{figure}

We studied reactions \Reacref{eq:t_co2}--\Reacref{eq:c2_hcooh} using the 18 \ce{H2O} cluster presented in \citet{Rimola2014, Perrero2022}. Subsequently, t-HOCO and c-HOCO are placed with random orientations on three different positions in search of distinct binding sites. In the case of level \one, only t-HOCO is placed at each binding site, as this level is exclusively used in the study of \Reacref{eq:t_co2}. By contrast, in the case of level \two, we study the adsorption of both c-HOCO and t-HOCO, as this is the model chemistry used for reactions \Reacref{eq:c_co2}-\Reacref{eq:c2_hcooh}. Once the three adsorption sites are obtained for each level and isomer, we start the sampling of the different reaction channels using relaxed PES scans. Transition states are later optimized and energy-refined from the maximum of these scans. Once the transition states are collected in all binding sites, pre-reactant complexes and product states are obtained by performing energy minimizations at both sides of the reaction coordinate at the TS. Activation energies ($\Delta U_{A}$) and reaction energies $\Delta U_{R}$ are computed as energy differences, Zero Point Vibrational Energy (ZPVE) inclusive, between the transition and reactant state, and reactant and product state, respectively. Reactions are categorized as ``barrierless'' if (a) the potential energy scan is clearly downhill; and (b) if, after inclusion of ZPVE contributions, the otherwise emerged barrier submerges. In the case of reactions \Reacref{eq:t2_hcooh} and \Reacref{eq:c2_hcooh} we used downhill intrinsic reaction coordinate calculations to prove the barrierless nature of the reaction (See Section \ref{sec:hcooh}). While we report the individual values for each reaction and binding site, we also provide the average of both quantities as the element of comparison between reactions. 

The energetic descriptors of the reactions (i.e., barriers and reaction energies) are subsequently used as inputs for a kinetic analysis based on transition state theory, incorporating an Eckart correction to account for quantum tunneling. While Eckart corrections provide only an approximation (specifically, a one-dimensional, zero-curvature correction \citep{nandi_quantum_2024}) the application of more sophisticated techniques is prohibitively expensive for these reactions. Additionally, the strong influence of the ice matrix on the reaction prevents us from reliably employing a gas-phase model to determine rate constants. 
Besides, all the hydrogenation reactions where tunneling plays a role, present a very high crossover temperature for tunneling, i.e., temperatures at which quantum effects start to dominate. This dramatically increases the cost of more sophisticated techniques, like instanton theory (see, e.g., \citet{kastner_theory_2014}).
We postpone the determination of more accurate rate constants for future works, focusing in the current investigation on a semi-quantitative description of the reactivity of HOCO.



\section{Results} \label{sec:results}

A summary of the energetic quantities for the hydrogenation of HOCO is gathered in \Tabref{tab:reactions_first}, and developed upon in the following sections.

\begin{table*}
	\begin{center}
	\caption{Reaction energies ($\Delta U_{R}$ in \kcalmol), activation energies ($\Delta U_{A}$ in \kcalmol), and transition state's imaginary frequency value ($\nu_i$ in cm$^{-1}$) for the reactions considered in this work, on the different binding sites. Barrierless reactions are labeled ``BL''.}
	\label{tab:reactions_first}
	\begin{tabular}{ccccccc}
	\toprule
	Reaction & Label & Computational Level & Site & $\Delta U_{R}$ & $\Delta U_{A}$ & $\nu_i$ \\
	\bottomrule
    \multirow{3}{*}{\ce{t-HOCO + H -> CO2 + H2}} & \multirow{3}{*}{\Reacref{eq:t_co2}} & \multirow{3}{*}{\one} & Site I & -86.5 & 16.8 & 3321i \\
    & & & Site II & -90.9 & 14.3 & 3234i \\
    & & & Site III & -92.0 & 12.6 & 3338i \\
    \midrule
    \textbf{Average $\Delta U_{R}$}: & \multicolumn{2}{c}{\bf -89.8} & 
    \textbf{Average $\Delta U_{A}$}: &  \multicolumn{2}{c}{\bf 14.6} \\
    \midrule
    \multirow{3}{*}{\ce{c-HOCO + H -> CO2 + H2}} & \multirow{3}{*}{\Reacref{eq:c_co2}} & \multirow{3}{*}{\two} & Site I & -94.9 & 0.1 & 585i \\
    & & & Site II & -93.6 & 1.3 & 510i \\
    & & & Site III & -94.8 & 0.4 & 408i \\
    \midrule
    \textbf{Average $\Delta U_{R}$}: & \multicolumn{2}{c}{\bf -94.5} & 
    \textbf{Average $\Delta U_{A}$}: & \multicolumn{2}{c}{\bf 0.6 } \\
    \midrule
    \multirow{3}{*}{\ce{t-HOCO + H -> CO + H2O}} & \multirow{3}{*}{\Reacref{eq:t_co}} & \multirow{3}{*}{\two} & Site I & -74.3 & BL & N/A \\
    & & & Site II & -73.5 & BL & N/A \\
    & & & Site III & -76.1  & BL & N/A \\
    \midrule
    \textbf{Average $\Delta U_{R}$}: & \multicolumn{2}{c}{\bf -74.6} & 
    \textbf{Average $\Delta U_{A}$}: & \multicolumn{2}{c}{\bf BL} \\
    \midrule   
    \multirow{3}{*}{\ce{c-HOCO + H -> CO + H2O}$^{a}$} & \multirow{3}{*}{\Reacref{eq:c_co}} & \multirow{3}{*}{\two} & Site I & -80.3 & <7.4 & 1425i \\
    & & & Site II & -76.6 & <8.9 & 1434i \\
    & & & Site III & -78.2 & <8.8 & 1390i \\
    \midrule
    \textbf{Average $\Delta U_{R}$}: & \multicolumn{2}{c}{\bf -78.2} & 
    \textbf{Average $\Delta U_{A}$}: & \multicolumn{2}{c}{\bf <8.4} \\
    \midrule
    \multirow{3}{*}{\ce{t-HOCO + H -> c-HCOOH}$^{b}$} & \multirow{3}{*}{\Reacref{eq:t2_hcooh}} & \multirow{3}{*}{\two} & Site I & -91.0 & BL & N/A \\
    & & & Site II & -89.5 & BL & N/A \\
    & & & Site III & -90.6 & BL & N/A \\
    \midrule
    \textbf{Average $\Delta U_{R}$}: & \multicolumn{2}{c}{\bf -90.4} & 
    \textbf{Average $\Delta U_{A}$}: & \multicolumn{2}{c}{\bf BL} \\
    \midrule
    \multirow{3}{*}{\ce{c-HOCO + H -> t-HCOOH}$^{b}$} & \multirow{3}{*}{\Reacref{eq:c2_hcooh}} & \multirow{3}{*}{\two} & Site I & -95.4 & BL & N/A \\
    & & & Site II & -104.5 & BL & N/A \\
    & & & Site III & -96.8 & BL & N/A \\
    \midrule
    \textbf{Average $\Delta U_{R}$}: & \multicolumn{2}{c}{\bf -98.9} & 
    \textbf{Average $\Delta U_{A}$}: & \multicolumn{2}{c}{\bf BL} \\
	\bottomrule
	\end{tabular}
    \tablefoot{$^{a}$- Constrained optimization (see \Secref{sec:co_h2o}). $^{b}$- Calculated from the bimolecular system (asymptote) as $\Delta U_{R}$ = $U_{\rm prod}$ - ($U_{\rm react}$ + $E_{\rm H}$)}
	\end{center}
	\end{table*}

\subsection{\ce{HOCO + H -> CO2 + H2} (Reactions \ref{eq:t_co2} and \ref{eq:c_co2})} \label{sec:co2}

Reactions \Eqref{eq:t_co2} and \Eqref{eq:c_co2} directly lead to \ce{CO2}. Our benchmark in the gas phase demonstrates that this reaction differs qualitatively between t-HOCO and c-HOCO. For t-HOCO, reaction \ref{eq:t_co2} exhibits the highest barrier identified in this study, measured at 12.2 \kcalmol using the XMS-CASPT2/cc-pVTZ energies as reference. Our benchmark analysis further reveals that these reactions are better described at the \one theoretical level, although DFT introduces an error (see Section \ref{sec:method:bs}). 

\begin{figure*}
    \centering
    \includegraphics[width=0.85\linewidth]{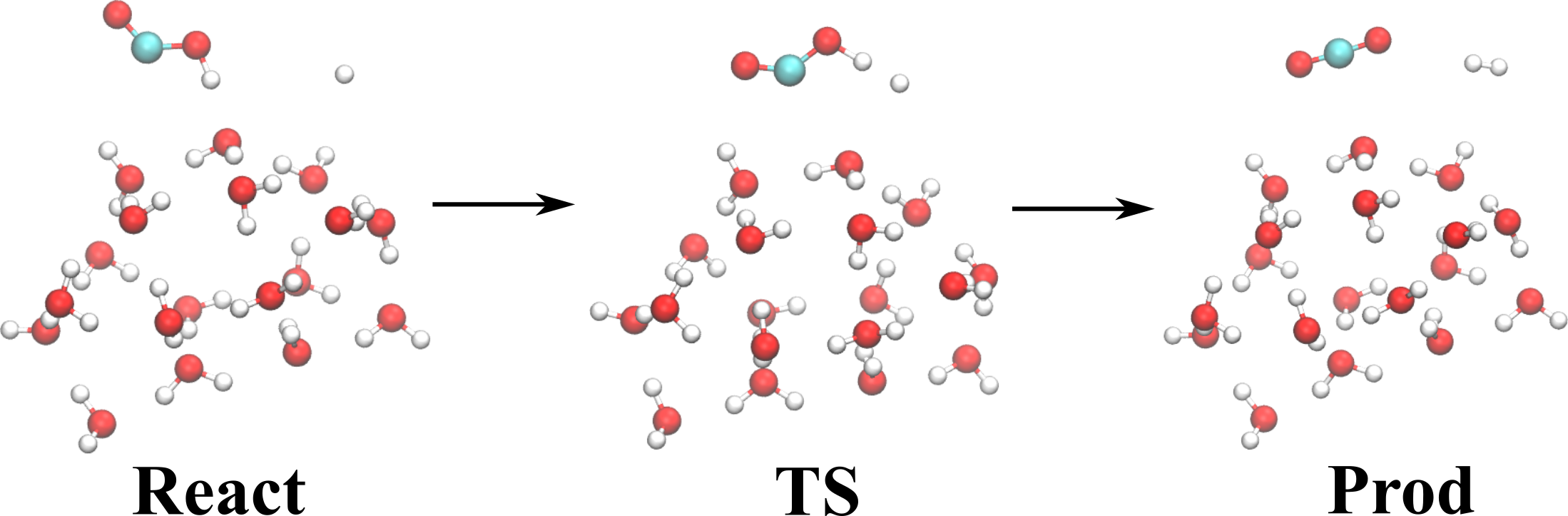}
    \caption{Snapshots of the stationary points for reaction \ref{eq:t_co2}. For simplicity, we only show a single binding site, site II in this case.}
    \label{fig:Structures_CO2}
\end{figure*}

The calculations on the ice cluster reveal a behavior similar to that observed in the gas phase. In Figure \ref{fig:Structures_CO2}, we provide a visual representation of this reaction. The averaged activation energy, $\Delta \overline{U}_{A, \Reacref{eq:t_co2}}$, across the three binding sites is 14.6 \kcalmol, which is higher than the gas-phase results discussed in \Secref{sec:method:bs}. The specific $\Delta \overline{U}_{A, \Reacref{eq:t_co2}}$ values for binding sites I, II, and III are 16.8, 14.3, and 12.6 \kcalmol, respectively, indicating a moderate influence of the binding site. This increase in $\Delta \overline{U}_{A, \Reacref{eq:t_co2}}$ is attributed to the additional energy required to cleave the hydrogen bond between the OH group in t-HOCO and the adjacent water molecule on the surface (see Figure \ref{fig:Structures_CO2} for an illustration of this effect). Moreover, \Reacref{eq:t_co2} exhibits an unusually high imaginary transition frequency, with an absolute value of approximately 3330 cm$^{-1}$, which remains consistent across binding sites. Such high transition frequencies suggest that quantum tunneling plays a central role in the reaction, making it impossible to dismiss \Reacref{eq:t_co2} solely based on energetic considerations. This aspect is explored further in \Secref{sec:kinetics}.

In contrast with \Reacref{eq:t_co2}, the same reaction in the \textit{cis} isomer, \Reacref{eq:c_co2}, presents a tiny $\Delta \overline{U}_{A, \Reacref{eq:c_co2}}$, of around 0.6 \kcalmol with limited variability with binding sites. The H-abstraction reaction presents similar exothermicity ($\Delta U_{R}$) for both \Reacref{eq:t_co2} and \Reacref{eq:c_co2}, -89.9 and -94.5 \kcalmol respectively. Focusing on $\Delta \overline{U}_{A}$, the striking difference of more than 15 \kcalmol between H-abstraction in c-HOCO and t-HOCO is an unconventional effect difficult to guess \textit{a priori}. More details on the reason behind this distinct chemistry are given in \Secref{sec:ration}. In addition to the distinct chemical behavior of c/t-HOCO, it is also interesting to investigate the effect of the ice matrix in this reaction. While our benchmark hints at \Reacref{eq:c_co2} being a barrierless reaction, a small but noticeable barrier emerges on the ice cluster. The $\Delta \overline{U}_{A}$ for \Reacref{eq:c_co2} remains very low not changing the qualitative picture for the reaction. It is important to emphasize the delicate balance of electronic effects in reactions occurring on interstellar ices. In these systems, an additional contribution to $\Delta U_{A}$ must be overcome, arising from the loss of stability due to breaking hydrogen bonds between the adsorbates and water.

\subsection{\ce{HOCO + H -> CO + H2O} (Reactions \ref{eq:t_co} and \ref{eq:c_co})} \label{sec:co_h2o}

\begin{figure*}
    \centering
    \includegraphics[width=0.9\linewidth]{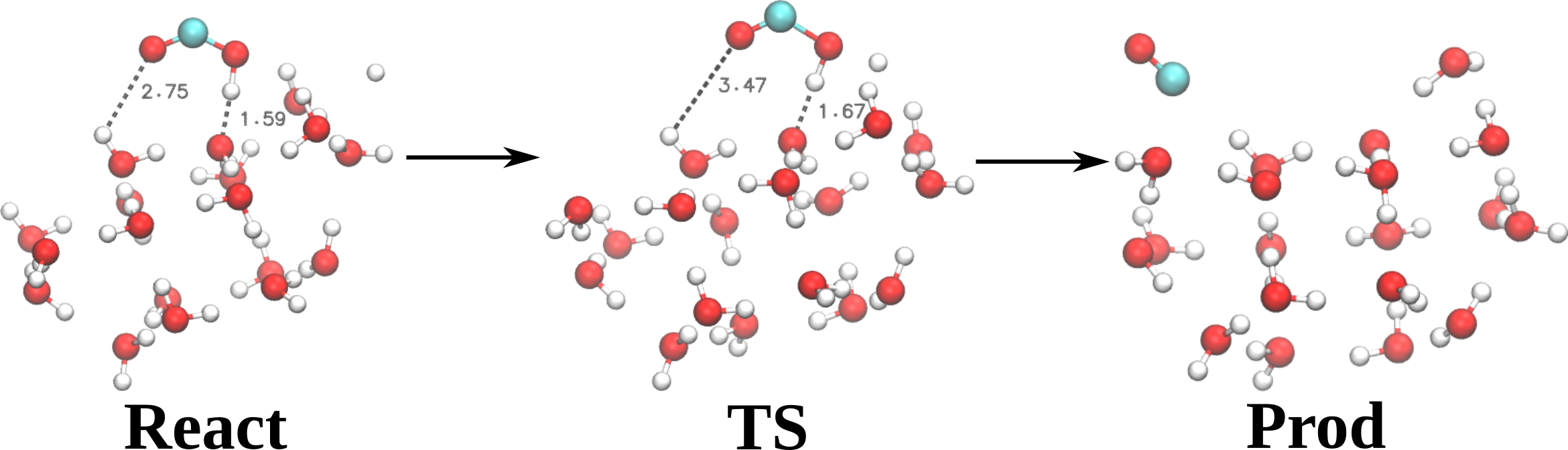}
    \caption{Snapshots of the stationary points for reaction \ref{eq:c_co}. Site I is shown in this case. Distances between c-HOCO and the surface are indicated to highlight the structural changes at the TS. }
    \label{fig:Structures_CO}
\end{figure*}

We also identify a distinct chemical behavior depending on the isomerism of the HOCO radical for Reactions \Reacref{eq:t_co} and \Reacref{eq:c_co}. In contrast to the behavior observed for \Reacref{eq:t_co2} and \Reacref{eq:c_co2}, this new set of reactions reveals that c-HOCO has activation energies with a TS depicted in Figure \ref{fig:Structures_CO}, whereas \Reacref{eq:t_co} proceeds without a barrier.\footnote{There is a small barrier at the revTPSSH(D3BJ)-gCP/def2-SVP level that vanishes upon correction with a 3$\zeta$ basis set and inclusion of ZPVE contributions.}
Specifically, \Reacref{eq:c_co} has an average activation energy of $\Delta \overline{U}_{A, \Reacref{eq:c_co}} = 8.4$ \kcalmol (individual values: $\Delta U_{A, \Reacref{eq:c_co}\textrm{I}} = 7.4$ \kcalmol, $\Delta U_{A, \Reacref{eq:c_co}\textrm{II}} = 8.9$ \kcalmol, $\Delta U_{A, \Reacref{eq:c_co}\textrm{III}} = 8.8$ \kcalmol). However, this value is considered an upper bound (see below). The reaction energies show minimal variation depending on the initial HOCO isomer, with slightly higher values observed for \Reacref{eq:c_co} ($\Delta \overline{U}_{R, \Reacref{eq:c_co}} = -78.2$ \kcalmol compared to $\Delta \overline{U}_{R, \Reacref{eq:t_co}} = -74.6$ \kcalmol). It is mandatory to indicate that the search for the transition state for \Reacref{eq:c_co} required imposing a geometric constraint in the dihedral angle in the HOCO skeleton. Otherwise the optimization ends in t-HOCO, i.e., the optimizer suggests a spontaneous torsion at the transition state. Because of this, in reality our postulated transition state is formally a second order saddle point rather than a true transition state and our derived $\Delta \overline{U}_{A, \Reacref{eq:c_co}}$ should be preemptively considered an upper bound of the real one. Nonetheless, we expect our barriers to be close to the actual one, as it is close to 5.3 kcal mol$^{-1}$ (see Figure \ref{fig:benchmark}) calculated in our benchmark or at the revTPSSh(D3BJ)/def2-TZVPPD level for geometry optimization and energies (Appendix \ref{sec:app1}), where we did not find any spontaneous torsion of the system. We deepen in our reasoning as to why we consider that assuming a second order saddle point for \Reacref{eq:c_co} is a better approximation to model \Reacref{eq:c_co} than assuming a spontaneous torsion to t-HOCO in Appendix \ref{sec:app1}. Obviously, this problem is not found in the case of \Reacref{eq:t_co} because the reaction is found to be barrierless.


\subsection{\ce{HOCO + H -> HCOOH} (Reactions \ref{eq:t2_hcooh} and \ref{eq:c2_hcooh})} \label{sec:hcooh}

\begin{figure}
    \centering
    \includegraphics[width=\linewidth]{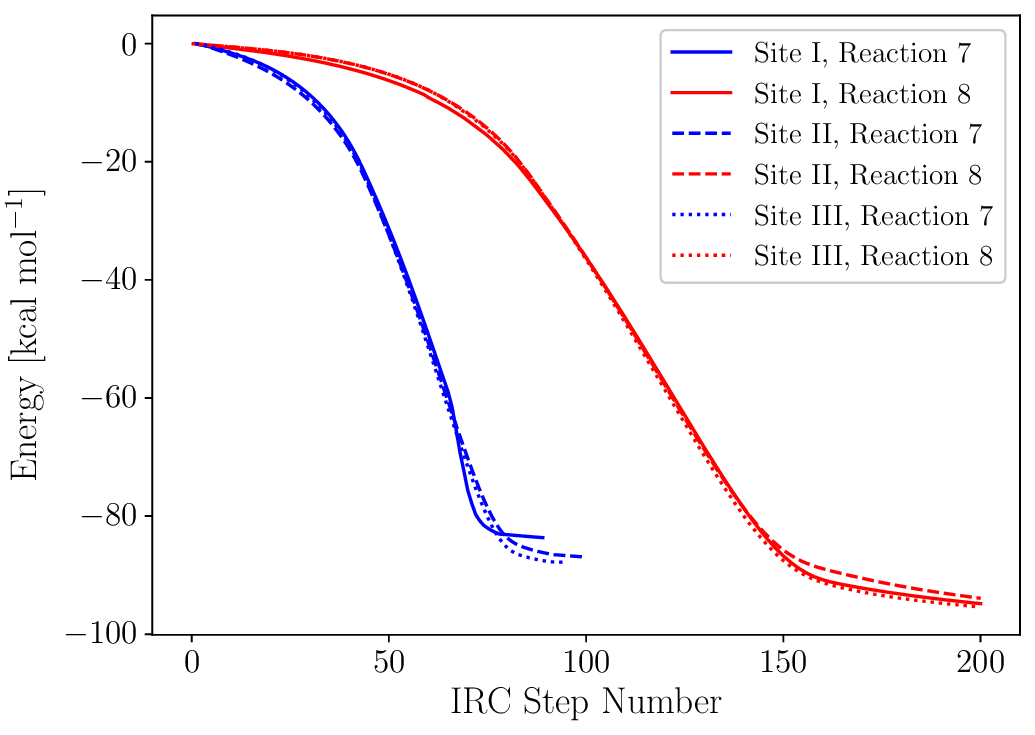}
    \caption{Downhill IRC profiles for reactions \Reacref{eq:t2_hcooh} and \Reacref{eq:c2_hcooh}. The profiles are calculated at the \two level of theory using a def2-SVP \citep{Weigend2005} basis set, i.e. without energy or ZPVE contributions.}
    \label{fig:irc}
\end{figure}

The final reaction considered is the formation of formic acid (\ce{HCOOH}) through hydrogenation at the carbon atom that formally carries the unpaired electron of the HOCO radical. This reaction is analyzed at the \two level of theory and is found to be barrierless. Due to challenges in properly converging the PES scans, we adopted an alternative approach. To demonstrate the absence of intrinsic barriers for this reaction on ices, we positioned an H-atom at an initial internuclear C–H distance of 3.00 \AA\xspace and conducted a downhill intrinsic reaction coordinate (dIRC) calculation. No energy refinement is performed for the dIRC, meaning that the calculations are done using the 2$\zeta$ basis set. All of our dIRC calculations unambiguously confirm the formation of \ce{HCOOH} (see dIRC profiles in Figure \ref{fig:irc}). Therefore, it is reasonable to assume that \Reacref{eq:t2_hcooh} and \Reacref{eq:c2_hcooh} are barrierless in astrochemical models.

\subsection{Kinetic analysis of the hydrogenation reactions} \label{sec:kinetics}

\begin{figure}
    \centering
    \includegraphics[width=\linewidth]{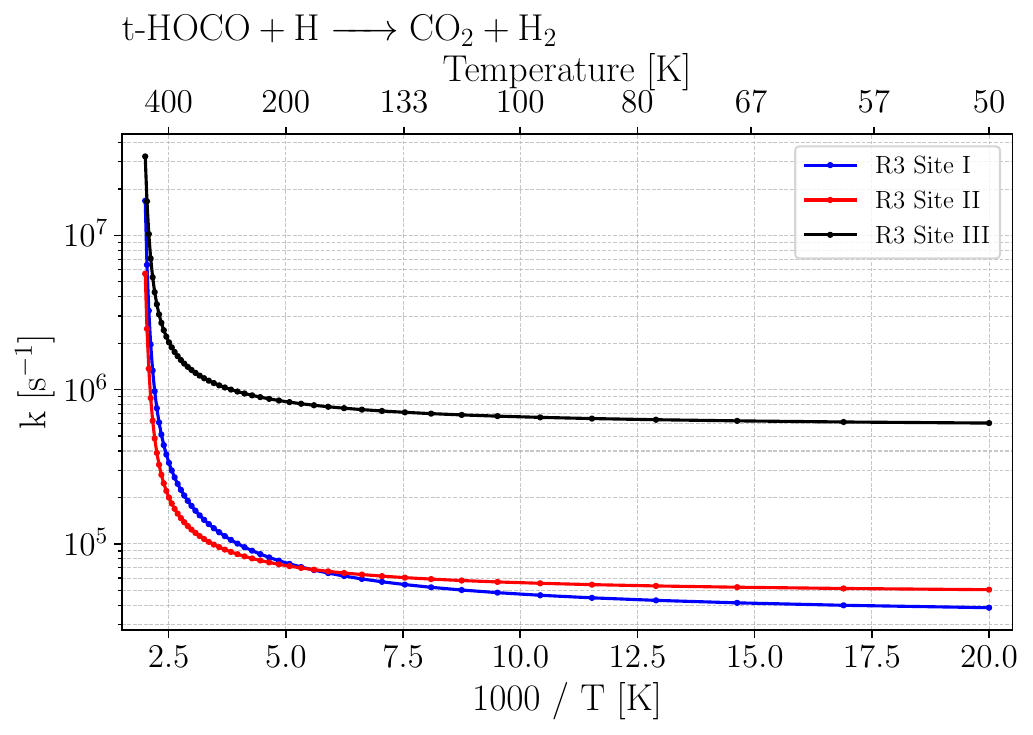} 
    \vfill
    \includegraphics[width=\linewidth]{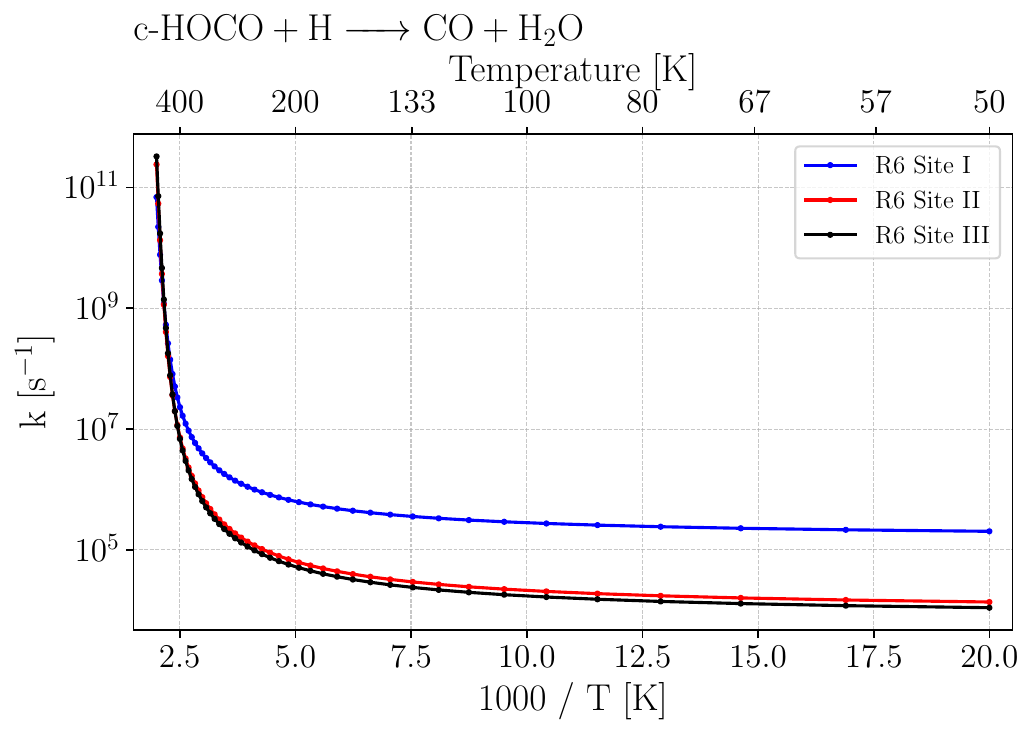} 
    \vfill
    \caption{Reaction rate constants for Reactions \Reacref{eq:t_co2} (top) and \Reacref{eq:c_co} (bottom) in different binding sites. }
    \label{fig:rates}
\end{figure}

In contrast to our previous work \citep{molpeceres_cracking_2023}, where microcanonical rate constants were derived, this study employs kinetic analysis for barrier-mediated reactions using conventional transition state theory. This approach is justified as all reactions considered here are elementary. The chosen code for these calculations was a developer version of DL-Find \citep{kae09a}. In all calculations, the rotational partition function is fixed to emulate the rigidity characteristic of a real ice slab. It is important to note that, although the TS and activation energies for \Reacref{eq:c_co} represent an upper bound and correspond to a second-order saddle point, the rate constants are derived using these parameters, hence making them a lower bound of the real ones (Appendix \ref{sec:app1}). Our analysis focuses on the rate constants for \Reacref{eq:t_co2} and \Reacref{eq:c_co}, as these reactions exhibit significant energy barriers. The remaining reactions are predominantly rapid due to their barrierless nature or very low $\Delta U_{A}$. For example, the lowest reaction rate constant observed for \Reacref{eq:c_co2} is 2.5$\times$10$^{10}$ s$^{-1}$. The rate constants for reactions \Reacref{eq:t_co2} and \Reacref{eq:c_co} using the upper bound values are plotted in Figure \ref{fig:rates} for temperatures down to 50 K, under the assumption that at lower temperatures the rate constants reach a horizontal asymptote. While this assumption is formally correct, it does not strictly apply to an Eckart treatment of quantum tunneling. However, Figure \ref{fig:rates} and the high crossover temperatures of \textgreater 700 K for \Reacref{eq:t_co2} and \textgreater 300 K for \Reacref{eq:c_co}, indicate that the rate constant's decay at low temperatures is small enough.

Both \Reacref{eq:t_co2} and \Reacref{eq:c_co} exhibit reaction rate constants ranging from 10$^{4}$ to 10$^{6}$ s$^{-1}$ at low temperatures (below $\sim$ 50 K). These values align with those reported by \citet{asg17} and \citet{SENEVIRATHNE201759} for the diffusion of H atoms, that is the competing process with reaction. In their studies on diffusion, the authors of both works report rate constants of around $10^{5}$--$10^{10}$ s$^{-1}$ at 20 K, and $10^{-2}$--$10^{10}$ s$^{-1}$ at 10 K, depending on the nature of the binding site and the ice phase (crystalline or amorphous). For amorphous solid water (ASW) on an average binding site \citet{SENEVIRATHNE201759} reports diffusion rate constants (See their Figure 7 bottom panels) of $\sim$$10^{4}$ s$^{-1}$, indicating that both \Reacref{eq:t_co2} and \Reacref{eq:c_co} could be viable in some cases despite their significant energy barriers. However, it is important to highlight the critical role of quantum tunneling in these processes, particularly for \Reacref{eq:t_co2}, which has a much higher crossover temperature and, consequently, greater uncertainty. The Eckart correction applied to the classical rate constants serves only as an approximation, providing an initial evaluation of reaction feasibility. Unfortunately, the calculated rate constants overlap with the range characteristic of hydrogen diffusion hopping rates, indicating an important diffusion-reaction competiton. This makes it difficult to definitively determine the importance of these reactions within the broader reaction network. Future iterations of this study will focus on deriving reaction rate constants using more advanced methods. It is a particularly challenging task, especially for \Reacref{eq:t_co2}, due to the complex electronic structure of the radicals (\Secref{sec:method:bs} and \Secref{sec:ration}), and the huge influence of quantum tunneling. This factor, added to the competition of \Reacref{eq:t_co2} with \Reacref{eq:t_co}, whose transition states are close makes us consider that we are likely overestimating $k$ for \Reacref{eq:t_co2} (Section \ref{sec:caveats}). For reaction \Reacref{eq:c_co}, we do not expect very high uncertainties, because the crossover temperature is not as elevated as in the previous case and because the reaction rate constants are a lower bound. Nevertheless, for the use of these reactions in chemical models, we recommend using the lowest reaction rate constants reported in this work for each reaction as a conservative estimate.

\subsection{\ce{c-HOCO -> t-HOCO} direct isomerization} \label{sec:isomers}

A legitimate question surrounding the hydrogenation of the HOCO radical or, in general, of any molecular isomer adsorbed on an interstellar surface, is whether the hydrogenation rate surpasses the rate for isomerization. To address this question, we compare the rate constant for the conversion from \ce{c-HOCO -> t-HOCO} (as the exothermic step) with the rate-limiting step for hydrogenation reactions in the ISM, i.e., the accretion of atomic hydrogen onto grain surfaces. According to \citet{Wakelam2017h2} the accretion rate of H on a dust grain is approximately 1.2$\times$10$^{-5}$ s$^{-1}$ (that corresponds to 1 atom day$^{-1}$), although the exact number depends on the steady state H abundance determined by the balance between the cosmic-ray ionization and \ce{H2} reformation \citep{Goldsmith2005}. Our results in the gas phase using CCSD(T)/aug-cc-pVTZ//revTPSSh(D3BJ)/def2-SVP yield $\Delta U_{A,iso}$=6.4 \kcalmol for \ce{c-HOCO -> t-HOCO}, and a reaction rate constant, $k_{\rm iso}$=3.47$\times$10$^{-5}$ s$^{-1}$ at 10 K, i.e. in the timescales of H accretion. We report the rate constants at 10 K for the isomerization reaction because the crossover temperature of isomerization in the gas phase is 144 K, which is low compared with those of \Reacref{eq:t_co2} and \Reacref{eq:c_co}. The crossover temperature is lower on the surface, translating in much lower rate constants, not competitive with hydrogenation. We investigated the same isomerization reaction on the ice surface using the DLPNO-CCSD(T)/jun-cc-pV(T+d)Z//revTPSSh(D3BJ)/def2-SVP method \citep{guo_communication_2018, papajak_perspectives_2011}. We found that $\Delta U_{A,iso}$ varies significantly depending on the binding site: {4.2, 6.7, 6.9} \kcalmol, for Site I,II and III, with Site I exhibiting stabilization of the TS through H-bonds with the surface. More importantly, the rate constants ($k_{\rm iso}$) for the reaction decrease in all cases to {6.0$\times$10$^{-15}$, 6.8$\times$10$^{-22}$, 9.1$\times$10$^{-21}$} s$^{-1}$. 

The decrease in rate constants between ice and gas is driven by the same factor observed in our study of thioformic acid formation \cite{Molpeceres2021b} through the OCSH radical (the sulfur equivalent of HOCO). On a surface, the molecular motion responsible for the isomerization is the migration of the CO moiety, rather than the OH (or SH) group. This motion involves a heavier group, leading to a reduction in the imaginary transition frequency, which in turn decreases quantum tunneling. Our estimated crossover temperatures for tunneling are below 85 K for the isomerization on the ice. Consequently, the \ce{c-HOCO -> t-HOCO} isomerization is much slower than hydrogenation, making the reactions discussed above (\Reacref{eq:t_co2} and \Reacref{eq:c_co}) more viable and realistic under ISM conditions. Finally, the back reaction \ce{t-HOCO -> c-HOCO} is endothermic to begin with, and therefore impossible in the cold ISM.

\section{Discussion} \label{sec:discussion}

\subsection{\ce{Chemical rationale}} \label{sec:ration}

\begin{figure}
	\centering
	\includegraphics[width=\linewidth]{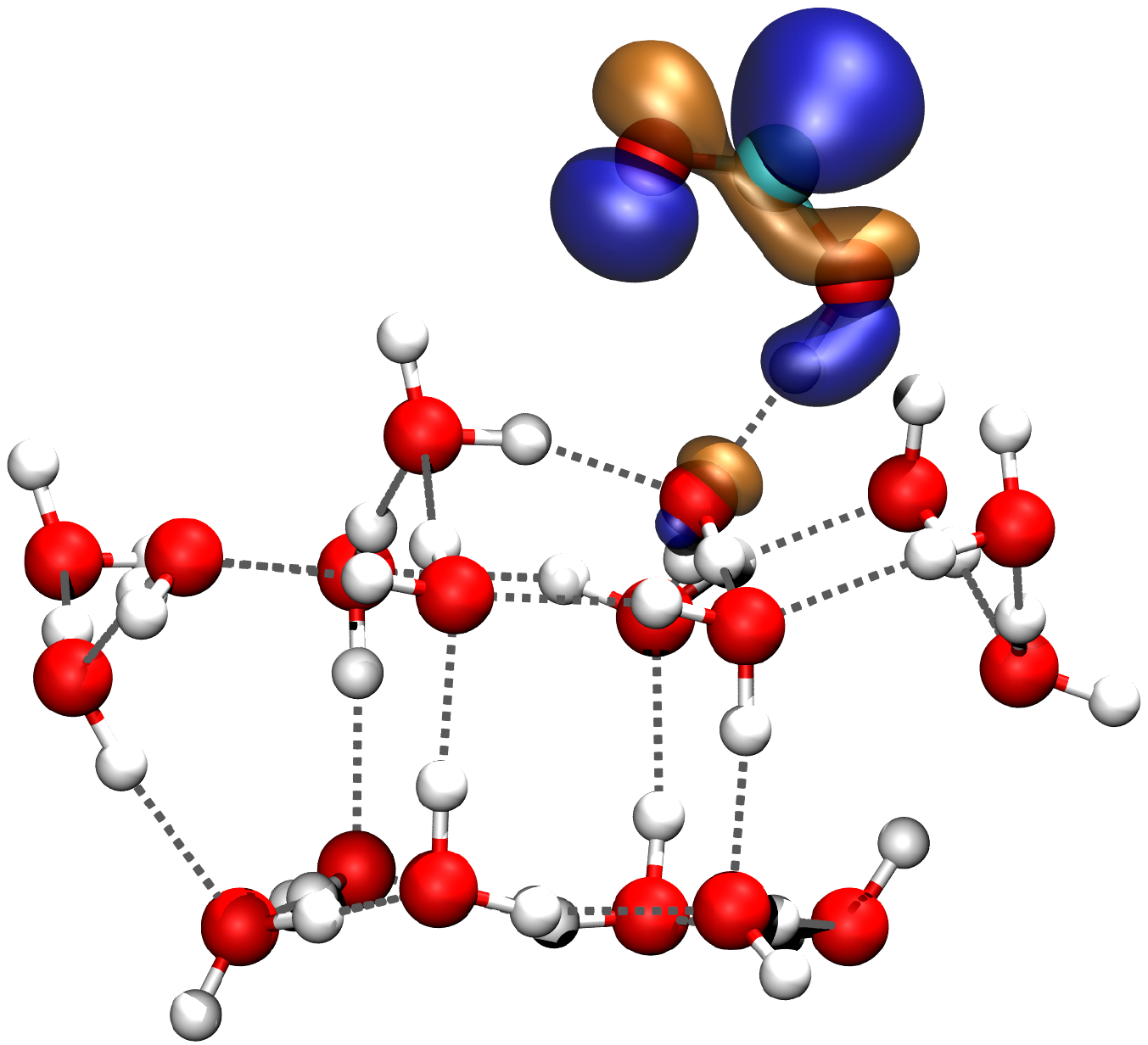}
	\vfill
    \vfill
	\includegraphics[width=\linewidth]{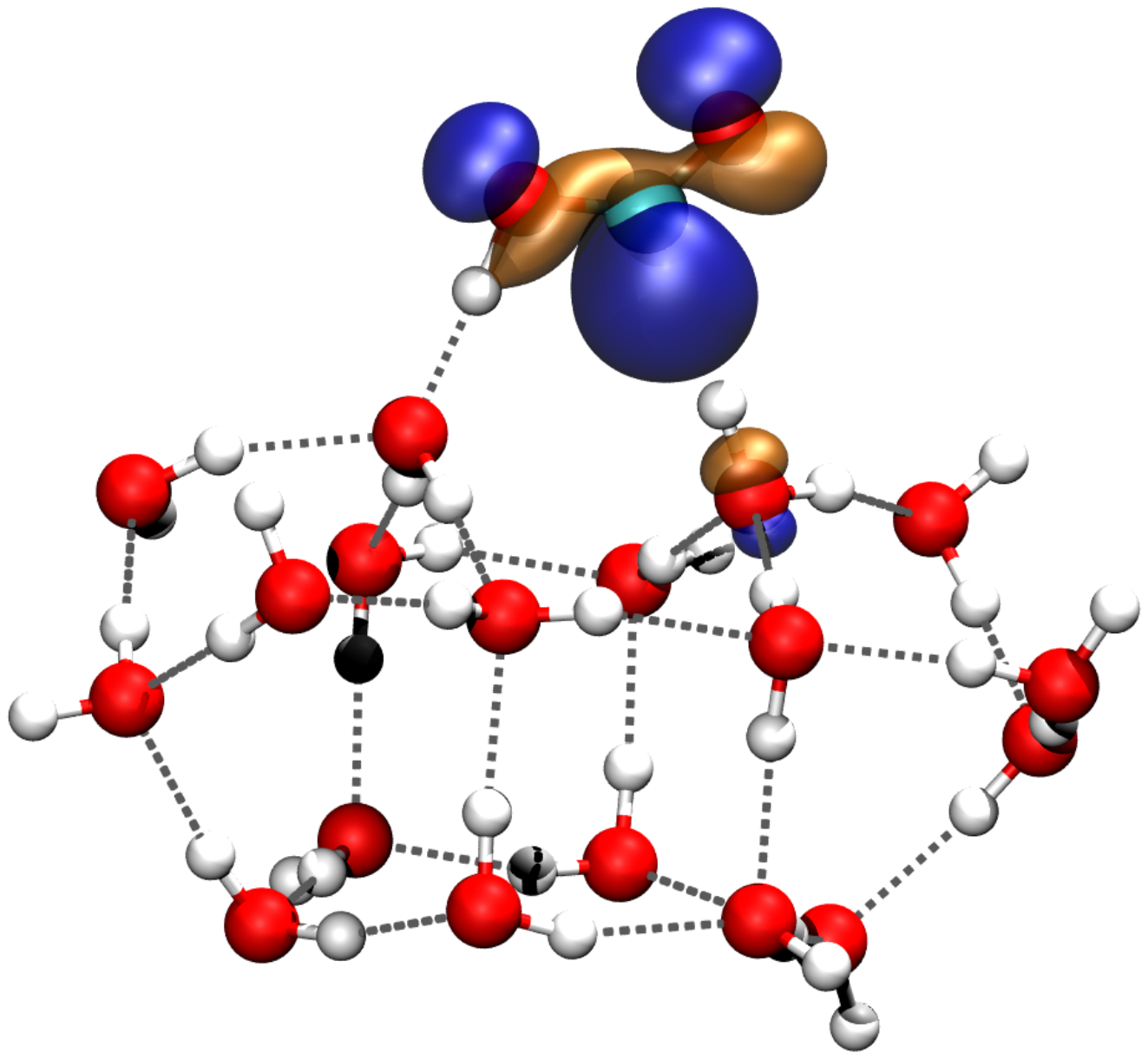}
	\caption{Singly occupied molecular orbitals (SOMOs) for the cis (top panel) and trans (bottom panel) conformers of HOCO on water ice cluster models. The orange and blue lobes represent the positive and negative regions of the wavefunction, respectively. The orbital plots are visualized with isovalues set to 0.05 a.u.}
	\label{fig:SOMOs}
\end{figure}

The key to understanding the counter-intuitive behavior of the radical-radical reaction between H and HOCO lies in the radical's electronic structure, specifically its frontier orbital, i.e., the singly occupied molecular orbital (SOMO). Figure \ref{fig:SOMOs} illustrates these orbitals for cis- and trans-HOCO on the water ice model, offering insight into how the electronic distribution shapes their reactivity.

For cis-HOCO, the SOMO extends over the O-H bond, promoting the barrierless H-abstraction reaction that yields \ce{CO2 + H2} (\Reacref{eq:c_co2}). However, the orbital does not extend to the oxygen’s lone pair, which explains the presence of an energy barrier for the path yielding \ce{CO + H2O} (\Reacref{eq:c_co}). In contrast, for trans-HOCO, the SOMO lacks electron density on the O-H bond but overlaps with the oxygen atom's lone pair. This configuration accounts for the barrierless formation of \ce{CO + H2O} (\Reacref{eq:t_co}) and the barrier-mediated formation of \ce{CO2 + H2} (\Reacref{eq:t_co2}). Notably, in both conformers, the unpaired electron in the SOMO is primarily localized on the carbon atom, as predicted by its Lewis structure. This characteristic also explains the formation of formic acid (\Reacref{eq:t2_hcooh} and \Reacref{eq:c2_hcooh}). These subtle orbital variations are able to explain the unconventional distinct reactivity found in these systems. The shape of the SOMO orbital on the ice resembles almost entirely the already reported SOMO in the gas phase \citep{mccarthy_isotopic_2016}.

\subsection{Astrophysical implications}

The formation of interstellar \ce{CO2} on the surface of interstellar ices is a topic of significant importance in modern astrochemistry due to the recent surge of ice observations using JWST \cite[e.g.,][]{McClure2023,rocha_jwst_2024, dartois_spectroscopic_2024}. For many years, the assumed reaction leading to \ce{CO2} was \Reacref{eq:co2} despite early evidences that such a reaction should not be efficient \citep{arasa_molecular_2013}. Very recently we managed to demonstrate, both theoretically \citep{molpeceres_cracking_2023} and experimentally \citep{ishibashi_proposed_2024} that indeed \Reacref{eq:co2} leads to the formation of HOCO, instead of \ce{CO2}. However, the presence of \ce{CO2} in traditional experiments for \ce{CO + OH}, \citep[e.g.][]{oba_experimental_2010,noble_co_2011,Qasim2019,ioppolo_surface_2011} strongly implies the presence of abstraction reactions, something that we already discussed on in our recent manuscripts. In this work, we identify an additional critical factor in the puzzle of \ce{CO2} formation. At very low temperatures, where (almost) only H atoms can diffuse and react, the isomeric form of the HOCO radical plays a crucial role in the formation of interstellar \ce{CO2} through hydrogenation. Determining the c-HOCO / t-HOCO branching ratio in \Reacref{eq:co2} is highly challenging due to the fast interconversion of the isomers between before thermalization, after which the rate constants in \Secref{sec:isomers} are applicable and slow. Later, thermalized HOCO radicals can remain dormant until a H atom diffuses nearby. Depending on the relative orientation of the two radicals, they can then react through the reactions studied in this work (\Reacref{eq:t_co2}–\Reacref{eq:c2_hcooh}):

\begin{align*}
    \ce{t-HOCO + H &-> CO2 + H2} \tag{\small \text{Possibly slow}} \\
    \ce{c-HOCO + H &-> CO2 + H2} \tag{\small \text{Fast}}\\
    \ce{t-HOCO + H &-> CO + H2O} \tag{\small \text{Fast}} \\
    \ce{c-HOCO + H &-> CO + H2O} \tag{\small \text{Competitive with H diffusion}}\\
    \ce{t-HOCO + H &-> c-HCOOH}  \tag{\small \text{Fast}} \\
    \ce{c-HOCO + H &-> t-HCOOH}  \tag{\small \text{Fast}}.
\end{align*}
\noindent The rate of formation of \ce{CO2} is therefore linked to the abundance of c-HOCO, as \ce{CO2} formation will be enhanced under conditions of excess c-HOCO. Since c-HOCO is less stable than t-HOCO by roughly 4 \kcalmol on ASW \citep{molpeceres_cracking_2023}, the prospect for \ce{CO2} formation through HOCO hydrogenation appears significantly less promising than previously assumed.
This is further aggravated by the fact that, as demonstrated in a recent work \citep{Molpeceres2022Thio}, c-HCOOH, the product of reaction \Reacref{eq:t2_hcooh}, reacts much faster for H-abstraction with another H (rate constants up to four orders of magnitude higher) than t-HCOOH, which is much less reactive.
This means that any c-HOCO molecule converted to t-HCOOH will remain as formic acid, rather than irreversibly convert to \ce{CO2 + H}. Given that t-HOCO transforms \ce{CO2} slowly or not at all (see below), the question arises whether HOCO is truly the dominant source of interstellar \ce{CO2} at low temperatures, and the \ce{CO + OH} reaction is a proxy for HOCO formation. The presence of \ce{CO2} in former experiments studying the reaction shows that it is indeed plausible. Therefore, the key question is not whether the reaction can occur, but if it serves as the dominant pathway in interstellar environments.
It is in this scenario where alternative reactions, like 

\begin{align}
    \ce{CO + O &-> CO2} \\
    \ce{HCO + O &-> CO2 + H},
\end{align}

\noindent take a protagonist role. Contemporary models to this article show that indeed the mobility of relatively heavy species like the oxygen atom is indeed possible when considering binding site heterogeneity \citep{furuya_framework_2024}. Besides, the mobility of the O atom was proved by experiments \citep{Minissale2016a}.  The \ce{CO + O} reaction is forbidden by spin, but experimental studies show its occurrence with a relatively low activation energy of $\sim$1.4 \kcalmol \citep{minissale_co_2013} although calculations found a larger barrier on carbonaceus substrates (5 \kcalmol; \cite{goumans_formation_2008}). Furthermore, the \ce{HCO + O -> CO2 + H} reaction is possibly barrierless \citep{goumans_formation_2008}, although confirmation would be needed. Overall, a critical reevaluation of potential \ce{CO2} formation pathways is essential to reconstruct the formation history of this key interstellar molecule. Understanding \ce{CO2}'s chemical origins is vital for constraining the carbon budget available for the synthesis of complex organic compounds, as \ce{CO2} serves as a stable sink within the carbon and oxygen reaction networks on ices \citep[see, for example,][]{suzuki_chemical_2024} for a recent modeling study.

In addition to the considerations on \ce{CO2} formation, the remarkable effect of the HOCO conformer on the reaction is worth discussing. Recent studies have emphasized the impact of isomerism on interstellar dust surfaces \citep{Molpeceres2021b, Molpeceres2022Thio}, supported by observational evidence of both equilibrium and non-equilibrium isomerism. For instance, some theoretical studies on imines and carboxylic acids provide compelling examples of this phenomenon \citep[see, e.g.,][]{GarciadelaConcepcion2021, GarciadelaConcepcion2022}. However, the finding that the chemistry of c-HOCO and t-HOCO is essentially opposite stands out, in our opinion, as one of the most surprising and counterintuitive results involving interstellar isomerism. A similar behavior was previously observed in hydrogen abstraction reactions of formic acid \citep{Molpeceres2022Thio}. 
In the construction of chemical reaction network, radical-radical reactions are usually assumed to have a significantly lower barrier, if any, than radical-neutral reactions. The present work provides an important counterexample to such an assumption.
It is unclear which isomeric species will have a distinct chemical behavior, and we expect to continue the search for new cases. Furthermore, to the best of our knowledge, an explicit inclusion of conformerism is currently absent in astrochemical models and such a task will be the subject of a future work. Lastly, HOCO (and hence HCOOH) is a promising precursor of carboxylic acids \citep{ishibashi_proposed_2024}. Given that apart from HCOOH, only \ce{CH3COOH} and \ce{OHCOOH} have been detected in space \citep{mehringer_detection_1997, sanz-novo_discovery_2023}, the chemistry of HOCO opens up an interesting acid-base chemistry in early stages of a molecular cloud worth investigating as follow up of this work.

\subsection{Caveats and necessary inputs} \label{sec:caveats}

While our calculations address the title reaction on a cluster with unprecedented detail, the complexity of the process prevents a complete understanding of its progression in interstellar environments. Two specific aspects remain unresolved and could benefit from further investigation, although addressing the associated technical and conceptual limitations will be highly challenging.

Firstly, as emphasized throughout this work, our kinetic analysis yields only approximate rate constants. Unfortunately, the predicted range for these constants, approximately 10$^{4-6}$ s$^{-1}$, coincides with the values typically expected for thermal hydrogen diffusion \citep{SENEVIRATHNE201759,asg17}. Consequently, we are unable to determine with as high accuracy as we would like whether \Reacref{eq:t_co2} and \Reacref{eq:c_co} occur on dust grains. In \Secref{sec:kinetics}, we have discussed that \Reacref{eq:t_co2} is likely slower than our predictions. This is based on the substantial activation energy we report and the abnormally high transition frequency (albeit consistent across binding sites) associated with this reaction (see \Tabref{tab:reactions_first}). Given the high activation energy, even a slight reduction in the transition frequency or a modest alteration in the barrier shape (effects not accounted for in our current tunneling implementation) could lead to rate constant variations by several orders of magnitude. That is the reason why we think we are overestimating the rate constants of \Reacref{eq:t_co2}.
Conversely, we believe that \Reacref{eq:c_co} may be underestimated, or at the very least fairly estimated. This assessment stems from the fact that the activation energy for \Reacref{eq:c_co} on ice could only be determined using a constrained optimization, as outlined in \Secref{sec:co_h2o} and Appendix \ref{sec:app1}. This approach introduces an repulsive term into the PES, artificially inflating the activation energy. A reduction in this activation energy would result in an increase in the reaction rate constants, thereby enhancing the viability of the reaction.

The second factor beyond the scope of the present work is the actual \ce{c-HOCO}/\ce{t-HOCO} ratio on interstellar grains. Assuming that no other significant source of \ce{HOCO} contributes beyond \Reacref{eq:co2}, the critical step determining the \ce{c-HOCO}/\ce{t-HOCO} ratio is the association of \ce{OH} and \ce{CO} molecules. In our previous study \citep{molpeceres_cracking_2023}, we found that the direct formation of \ce{c-HOCO} encounters a sizable barrier on \ce{H2O} ice and is not feasible on CO ice, at least thermally. Consequently, all \ce{CO + OH} interactions must mostly lead to \ce{t-HOCO}, which forms with excess energy capable of driving back-and-forth isomerizations prior to thermalization on the surface.
Given that \ce{t-HOCO} is more stable, it is likely to dominate post-thermalization, resulting in a \ce{c-HOCO}/\ce{t-HOCO} ratio of less than 1.0. However, determining the precise ratio requires specialized calculations that account for dynamic energy dissipation. Molecular dynamics simulations represent the most promising approach to address this challenge.

\section{Conclusions} \label{sec:conclusions}

In the light of what has been presented in this work, we conclude that, despite the prior assumption that every reaction channel in  \Reacref{eq:naive_abstraction} should proceed without or with a meager barrier, we found completely otherwise. In this work we outline a reaction, or rather a set of reactions, significantly more complex than previously anticipated, which leads us to extract the following conclusions from our work:

\begin{enumerate}
    \item The \ce{HOCO} radical isomerism has a crucial impact on the reaction outcomes. \ce{H2O} formation is favored in the case of \ce{t-HOCO} (\Reacref{eq:t_co}), while \ce{c-HOCO} (\Reacref{eq:c_co2}) promotes \ce{CO2} formation. However, the formation of \ce{t/c-HCOOH} is unaffected by the isomeric form of \ce{HOCO}.
    \item As presented in previous works \citep{Molpeceres2022Thio}, the H-abstraction on HCOOH does depend on the isomeric form of HCOOH, with c-HCOOH capable of reforming t-HOCO, which in turn affects the overall hydrogenation network.
    \item The direct isomerization of \ce{c-HOCO -> t-HOCO} likely will not occur on interstellar ices.
    \item Our kinetic analyses show that, in spite of the high barriers found for reactions the rate coefficient of \Reacref{eq:t_co2} and \Reacref{eq:c_co}, both reactions might happen thanks to quantum tunneling. However, we caution that \Reacref{eq:t_co2} may be significantly overestimated, while \Reacref{eq:c_co} may be underestimated. A more sophisticated treatment of quantum tunneling is necessary to completely rule out or confirm these reactions. Nevertheless, it seems clear that the efficiency of the reactions is significantly lower than what was expected for a radical-radical recombination.
    \item Based on the findings above, it becomes clear that a reevaluation of the dominant formation routes of interstellar \ce{CO2} is required. Additionally, there needs to be a reassessment of the expected abundance of other molecules, such as formic acid.
    \item The \ce{c-HOCO/t-HOCO} ratio on ices remains poorly constrained, which is a crucial parameter for maximizing the insights derived from the results presented here.
    \item From a conceptual standpoint, the significant influence that the isomeric form of \ce{HOCO} exerts on its subsequent chemistry suggests that isomerism plays a more prominent role in the chemical evolution of the ISM than previously anticipated. 
\end{enumerate}

Building on the findings of this work, future efforts could focus on refining our understanding of the role of quantum tunneling in the H-abstraction reaction, as well as conducting molecular dynamics studies to explore the long-term consequences of the \ce{CO + OH} reaction, i.e t-HOCO / c-HOCO ratio.  Additionally, exploring differences in reaction outcomes when using alternative abstracting radicals (e.g., \ce{NH2}, \ce{CH3}) could provide valuable insights, as prior studies such as \citet{ishibashi_proposed_2024} highlighted efficient H-abstraction processes from OH radicals. Ultimately, comprehensive astrochemical modeling will remain a crucial tool for assessing the impact of the derived parameters in interstellar environments. We plan to pursue the development of such models in the near future.

\section*{Acknowledgements}

G.M acknowledges the support of the grant RYC2022-035442-I funded by MICIU/AEI/10.13039/501100011033 and ESF+. G.M. also received support from project 20245AT016 (Proyectos Intramurales CSIC). We acknowledge the computational resources provided by bwHPC and the German Research Foundation (DFG) through grant no INST 40/575-1 FUGG (JUSTUS 2 cluster), the DRAGO computer cluster managed by SGAI-CSIC, and the Galician Supercomputing Center (CESGA). The supercomputer FinisTerrae III and its permanent data storage system have been funded by the Spanish Ministry of Science and Innovation, the Galician Government and the European Regional Development Fund (ERDF). Early parts of this work (GM) were also funded by Japan Society for the Promotion of Science (JSPS International Fellow P22013, and Grant-in-aid 22F22013).
J.E.R acknowledges the support of the Horizon Europe
Framework Programme (HORIZON) under the Marie
Skłodowska-Curie grant agreement No 101149067, ``ICE-CN”.
N.W and Y.A acknowledge the support by JSPS KAKENHI grant No. 20H05847. A.I. acknowledges the support by JSPS Grant-in-Aid for Scientific Research JP24K17107. 

\section*{Data Availability}

The data supporting this article will be shared upon reasonable request to any of the corresponding authors. The molecular structures supporting our calculations will be available on a Zenodo repository prior to the definitive publishing of the manuscript.



\bibliographystyle{mnras}
\bibliography{example} 

\begin{thebibliography}{}
\makeatletter
\relax
\def\mn@urlcharsother{\let\do\@makeother \do\$\do\&\do\#\do\^\do\_\do\%\do\~}
\def\mn@doi{\begingroup\mn@urlcharsother \@ifnextchar [ {\mn@doi@}
  {\mn@doi@[]}}
\def\mn@doi@[#1]#2{\def\@tempa{#1}\ifx\@tempa\@empty \href
  {http://dx.doi.org/#2} {doi:#2}\else \href {http://dx.doi.org/#2} {#1}\fi
  \endgroup}
\def\mn@eprint#1#2{\mn@eprint@#1:#2::\@nil}
\def\mn@eprint@arXiv#1{\href {http://arxiv.org/abs/#1} {{\tt arXiv:#1}}}
\def\mn@eprint@dblp#1{\href {http://dblp.uni-trier.de/rec/bibtex/#1.xml}
  {dblp:#1}}
\def\mn@eprint@#1:#2:#3:#4\@nil{\def\@tempa {#1}\def\@tempb {#2}\def\@tempc
  {#3}\ifx \@tempc \@empty \let \@tempc \@tempb \let \@tempb \@tempa \fi \ifx
  \@tempb \@empty \def\@tempb {arXiv}\fi \@ifundefined
  {mn@eprint@\@tempb}{\@tempb:\@tempc}{\expandafter \expandafter \csname
  mn@eprint@\@tempb\endcsname \expandafter{\@tempc}}}

\bibitem[\protect\citeauthoryear{Andersson, Malmqvist, Roos, Sadlej  \&
  Wolinski}{Andersson et~al.}{1990}]{andersson_second-order_1990}
Andersson K.,  Malmqvist P.~A.,  Roos B.~O.,  Sadlej A.~J.,   Wolinski K.,
  1990, \mn@doi [J. Phys. Chem.] {10.1021/j100377a012}, 94, 5483

\bibitem[\protect\citeauthoryear{Aquilante et~al.,}{Aquilante
  et~al.}{2020}]{aquilante_modern_2020}
Aquilante F.,  et~al., 2020, \mn@doi [The Journal of Chemical Physics]
  {10.1063/5.0004835}, 152, 214117

\bibitem[\protect\citeauthoryear{Arasa, Van~Hemert, Van~Dishoeck  \&
  Kroes}{Arasa et~al.}{2013}]{arasa_molecular_2013}
Arasa C.,  Van~Hemert M.~C.,  Van~Dishoeck E.~F.,   Kroes G.~J.,  2013, \mn@doi
  [The Journal of Physical Chemistry A] {10.1021/jp400065v}, 117, 7064

\bibitem[\protect\citeauthoryear{Asgeirsson, Jónsson  \& Wikfeldt}{Asgeirsson
  et~al.}{2017}]{asg17}
Asgeirsson V.,  Jónsson H.,   Wikfeldt K.~T.,  2017, \mn@doi [Journal of
  Physical Chemistry C] {10.1021/acs.jpcc.6b10636}, 121, 1648

\bibitem[\protect\citeauthoryear{Becke}{Becke}{1993}]{Becke1993}
Becke A.~D.,  1993, \mn@doi [J. Chem. Phys.] {10.1063/1.464913}, 98, 5648

\bibitem[\protect\citeauthoryear{Caracciolo et~al.,}{Caracciolo
  et~al.}{2018}]{caracciolo_combined_2018}
Caracciolo A.,  et~al., 2018, \mn@doi [The Journal of Physical Chemistry
  Letters] {10.1021/acs.jpclett.7b03439}, 9, 1229

\bibitem[\protect\citeauthoryear{Clément et~al.,}{Clément
  et~al.}{2023}]{clement_astrochemical_2023}
Clément A.,  et~al., 2023, \mn@doi [Astronomy \& Astrophysics]
  {10.1051/0004-6361/202346188}, 675, A165

\bibitem[\protect\citeauthoryear{Dartois et~al.,}{Dartois
  et~al.}{2024}]{dartois_spectroscopic_2024}
Dartois E.,  et~al., 2024, \mn@doi [Nature Astronomy]
  {10.1038/s41550-023-02155-x}

\bibitem[\protect\citeauthoryear{Dunning}{Dunning}{1989}]{Duning1989}
Dunning T.~H.,  1989, J. Chem. Phys., 90

\bibitem[\protect\citeauthoryear{Enrique-Romero et~al.,}{Enrique-Romero
  et~al.}{2020}]{enrique-romero_revisiting_2020}
Enrique-Romero J.,  et~al., 2020, \mn@doi [Monthly Notices of the Royal
  Astronomical Society] {10.1093/mnras/staa484}, 493, 2523

\bibitem[\protect\citeauthoryear{Fdez.~Galván et~al.,}{Fdez.~Galván
  et~al.}{2019}]{fdez_galvan_openmolcas_2019}
Fdez.~Galván I.,  et~al., 2019, \mn@doi [Journal of Chemical Theory and
  Computation] {10.1021/acs.jctc.9b00532}, 15, 5925

\bibitem[\protect\citeauthoryear{Francisco, Muckerman  \& Yu}{Francisco
  et~al.}{2010}]{francisco_hoco_2010}
Francisco J.~S.,  Muckerman J.~T.,   Yu H.-G.,  2010, \mn@doi [Accounts of
  Chemical Research] {10.1021/ar100087v}, 43, 1519

\bibitem[\protect\citeauthoryear{Frost, Sharkey  \& Smith}{Frost
  et~al.}{1991}]{Frost1991}
Frost M.~J.,  Sharkey P.,   Smith I.~W.,  1991, \mn@doi [Faraday Discussions of
  the Chemical Society] {10.1039/DC9919100305}, 91, 305

\bibitem[\protect\citeauthoryear{Furuya}{Furuya}{2024}]{furuya_framework_2024}
Furuya K.,  2024, \mn@doi [The Astrophysical Journal]
  {10.3847/1538-4357/ad6b20}, 974, 115

\bibitem[\protect\citeauthoryear{García de~la Concepción, Jiménez-Serra,
  Carlos~Corchado, Rivilla  \& Martín-Pintado}{García de~la Concepción
  et~al.}{2021}]{GarciadelaConcepcion2021}
García de~la Concepción J.,  Jiménez-Serra I.,  Carlos~Corchado J.,  Rivilla
  V.~M.,   Martín-Pintado J.,  2021, \mn@doi [The Astrophysical Journal
  Letters] {10.3847/2041-8213/abf650}, 912, L6

\bibitem[\protect\citeauthoryear{García de~la Concepción et~al.,}{García
  de~la Concepción et~al.}{2022}]{GarciadelaConcepcion2022}
García de~la Concepción J.,  et~al., 2022, \mn@doi [Astronomy \&
  Astrophysics] {10.1051/0004-6361/202142287}, 658, A150

\bibitem[\protect\citeauthoryear{Garrod \& Pauly}{Garrod \&
  Pauly}{2011}]{Garrod2011}
Garrod R.~T.,  Pauly T.,  2011, \mn@doi [Astrophysical Journal]
  {10.1088/0004-637X/735/1/15}, 735, 15

\bibitem[\protect\citeauthoryear{{Goldsmith} \& {Li}}{{Goldsmith} \&
  {Li}}{2005}]{Goldsmith2005}
{Goldsmith} P.~F.,  {Li} D.,  2005, \mn@doi [Astrophys. J.] {10.1086/428032},
  \href {https://ui.adsabs.harvard.edu/abs/2005ApJ...622..938G} {622, 938}

\bibitem[\protect\citeauthoryear{Goumans, Uppal  \& Brown}{Goumans
  et~al.}{2008}]{goumans_formation_2008}
Goumans T. P.~M.,  Uppal M.~A.,   Brown W.~A.,  2008, \mn@doi [MNRAS]
  {10.1111/j.1365-2966.2007.12788.x}, 384, 1158

\bibitem[\protect\citeauthoryear{Granovsky}{Granovsky}{2011}]{granovsky_extended_2011}
Granovsky A.~A.,  2011, \mn@doi [J. Chem. Phys.] {10.1063/1.3596699}, 134,
  214113

\bibitem[\protect\citeauthoryear{Greenblatt \& Howard}{Greenblatt \&
  Howard}{1989}]{greenblatt_oxygen_1989}
Greenblatt G.~D.,  Howard C.~J.,  1989, \mn@doi [The Journal of Physical
  Chemistry] {10.1021/j100340a006}, 93, 1035

\bibitem[\protect\citeauthoryear{Grimme, Antony, Ehrlich  \& Krieg}{Grimme
  et~al.}{2010}]{grimme2010}
Grimme S.,  Antony J.,  Ehrlich S.,   Krieg H.,  2010, \mn@doi [Journal of
  Chemical Physics] {10.1063/1.3382344}, 132, 154104

\bibitem[\protect\citeauthoryear{Grimme, Ehrlich  \& Goerigk}{Grimme
  et~al.}{2011}]{Grimme2011}
Grimme S.,  Ehrlich S.,   Goerigk L.,  2011, \mn@doi [Journal of Computational
  Chemistry] {https://doi.org/10.1002/jcc.21759}, 32, 1456

\bibitem[\protect\citeauthoryear{Guo, Riplinger, Becker, Liakos, Minenkov,
  Cavallo  \& Neese}{Guo et~al.}{2018}]{guo_communication_2018}
Guo Y.,  Riplinger C.,  Becker U.,  Liakos D.~G.,  Minenkov Y.,  Cavallo L.,
  Neese F.,  2018, \mn@doi [The Journal of Chemical Physics]
  {10.1063/1.5011798}, 148, 011101

\bibitem[\protect\citeauthoryear{Ioppolo, van Boheemen, Cuppen, van Dishoeck
  \& Linnartz}{Ioppolo et~al.}{2011}]{ioppolo_surface_2011}
Ioppolo S.,  van Boheemen Y.,  Cuppen H.~M.,  van Dishoeck E.~F.,   Linnartz
  H.,  2011, \mn@doi [Monthly Notices of the Royal Astronomical Society]
  {10.1111/j.1365-2966.2011.18306.x}, 413, 2281

\bibitem[\protect\citeauthoryear{Ishibashi, Molpeceres, Hidaka, Oba, Lamberts
  \& Watanabe}{Ishibashi et~al.}{2024}]{ishibashi_proposed_2024}
Ishibashi A.,  Molpeceres G.,  Hidaka H.,  Oba Y.,  Lamberts T.,   Watanabe N.,
   2024, \mn@doi [The Astrophysical Journal] {10.3847/1538-4357/ad8235}, 976,
  162

\bibitem[\protect\citeauthoryear{K\"astner, Carr, Keal, Thiel, Wander  \&
  Sherwood}{K\"astner et~al.}{2009}]{kae09a}
K\"astner J.,  Carr J.~M.,  Keal T.~W.,  Thiel W.,  Wander A.,   Sherwood P.,
  2009, \mn@doi [J. Phys. Chem. A] {10.1021/jp9028968}, 113, 11856

\bibitem[\protect\citeauthoryear{Kästner}{Kästner}{2014}]{kastner_theory_2014}
Kästner J.,  2014, \mn@doi [Wiley Interdisciplinary Reviews: Computational
  Molecular Science] {10.1002/wcms.1165}, 4, 158

\bibitem[\protect\citeauthoryear{Lakin, Troya, Schatz  \& Harding}{Lakin
  et~al.}{2003}]{lakin_quasiclassical_2003}
Lakin M.~J.,  Troya D.,  Schatz G.~C.,   Harding L.~B.,  2003, \mn@doi [J.
  Chem. Phys.] {10.1063/1.1602061}, 119, 5848

\bibitem[\protect\citeauthoryear{Li, Chen, Zhang  \& Guo}{Li
  et~al.}{2014}]{li_quantum_2014}
Li J.,  Chen J.,  Zhang D.~H.,   Guo H.,  2014, \mn@doi [The Journal of
  Chemical Physics] {10.1063/1.4863138}, 140, 044327

\bibitem[\protect\citeauthoryear{Ma, Li  \& Guo}{Ma et~al.}{2012}]{Ma2012}
Ma J.,  Li J.,   Guo H.,  2012, \mn@doi [The Journal of Physical Chemistry
  Letters] {10.1021/jz301064w}, 3, 2482

\bibitem[\protect\citeauthoryear{Masunov, Wait  \& Vasu}{Masunov
  et~al.}{2018}]{masunov_catalytic_2018}
Masunov A.~E.,  Wait E.~E.,   Vasu S.~S.,  2018, \mn@doi [The Journal of
  Physical Chemistry A] {10.1021/acs.jpca.8b04501}, 122, 6355

\bibitem[\protect\citeauthoryear{McCarthy, Martinez, McGuire, Crabtree,
  Martin-Drumel  \& Stanton}{McCarthy et~al.}{2016}]{mccarthy_isotopic_2016}
McCarthy M.~C.,  Martinez O.,  McGuire B.~A.,  Crabtree K.~N.,  Martin-Drumel
  M.-A.,   Stanton J.~F.,  2016, \mn@doi [The Journal of Chemical Physics]
  {10.1063/1.4944070}, 144, 124304

\bibitem[\protect\citeauthoryear{{McClure} et~al.,}{{McClure}
  et~al.}{2023}]{McClure2023}
{McClure} M.~K.,  et~al., 2023, \mn@doi [Nature Astronomy]
  {10.1038/s41550-022-01875-w}, \href
  {https://ui.adsabs.harvard.edu/abs/2023NatAs.tmp...25M} {}

\bibitem[\protect\citeauthoryear{Mehringer, Snyder, Miao  \& Lovas}{Mehringer
  et~al.}{1997}]{mehringer_detection_1997}
Mehringer D.~M.,  Snyder L.~E.,  Miao Y.,   Lovas F.~J.,  1997, \mn@doi [The
  Astrophysical Journal] {10.1086/310612}, 480, L71

\bibitem[\protect\citeauthoryear{Minissale, Congiu, Manicò, Pirronello  \&
  Dulieu}{Minissale et~al.}{2013}]{minissale_co_2013}
Minissale M.,  Congiu E.,  Manicò G.,  Pirronello V.,   Dulieu F.,  2013,
  \mn@doi [Astronomy \& Astrophysics] {10.1051/0004-6361/201321453}, 559, A49

\bibitem[\protect\citeauthoryear{Minissale, Congiu  \& Dulieu}{Minissale
  et~al.}{2016}]{Minissale2016a}
Minissale M.,  Congiu E.,   Dulieu F.,  2016, \mn@doi [Astronomy and
  Astrophysics] {10.1051/0004-6361/201526702}, 585

\bibitem[\protect\citeauthoryear{{Molpeceres, G.} et~al.,}{{Molpeceres, G.}
  et~al.}{2022}]{Molpeceres2022Thio}
{Molpeceres, G.} et~al., 2022, \mn@doi [A\&A] {10.1051/0004-6361/202243366},
  663, A41

\bibitem[\protect\citeauthoryear{Molpeceres, Kästner, Fedoseev, Qasim,
  Schömig, Linnartz  \& Lamberts}{Molpeceres
  et~al.}{2021a}]{Molpeceres2021carbon}
Molpeceres G.,  Kästner J.,  Fedoseev G.,  Qasim D.,  Schömig R.,  Linnartz
  H.,   Lamberts T.,  2021a, \mn@doi [Journal of Physical Chemistry Letters]
  {10.1021/acs.jpclett.1c02760}, 12, 10854

\bibitem[\protect\citeauthoryear{Molpeceres, García de~la Concepción  \&
  Jiménez-Serra}{Molpeceres et~al.}{2021b}]{Molpeceres2021b}
Molpeceres G.,  García de~la Concepción J.,   Jiménez-Serra I.,  2021b,
  \mn@doi [The Astrophysical Journal] {10.3847/1538-4357/ac2ebc}, 923, 159

\bibitem[\protect\citeauthoryear{Molpeceres, Enrique-Romero  \&
  Aikawa}{Molpeceres et~al.}{2023}]{molpeceres_cracking_2023}
Molpeceres G.,  Enrique-Romero J.,   Aikawa Y.,  2023, \mn@doi [Astronomy \&
  Astrophysics] {10.1051/0004-6361/202347097}, 677, A39

\bibitem[\protect\citeauthoryear{Molpeceres, Tsuge, Furuya, Watanabe,
  San~Andrés, Rivilla, Colzi  \& Aikawa}{Molpeceres
  et~al.}{2024}]{molpeceres_carbon_2024}
Molpeceres G.,  Tsuge M.,  Furuya K.,  Watanabe N.,  San~Andrés D.,  Rivilla
  V.~M.,  Colzi L.,   Aikawa Y.,  2024, \mn@doi [The Journal of Physical
  Chemistry A] {10.1021/acs.jpca.3c08286}, p. acs.jpca.3c08286

\bibitem[\protect\citeauthoryear{Nandi, Molpeceres, Gupta, Major, Kästner,
  Martin  \& Kozuch}{Nandi et~al.}{2024}]{nandi_quantum_2024}
Nandi A.,  Molpeceres G.,  Gupta P.~K.,  Major D.~T.,  Kästner J.,  Martin
  J.~M.,   Kozuch S.,  2024, in , Comprehensive {Computational} {Chemistry}.
Elsevier, pp 713--734, \mn@doi{10.1016/B978-0-12-821978-2.00020-9}, \url
  {https://linkinghub.elsevier.com/retrieve/pii/B9780128219782000209}

\bibitem[\protect\citeauthoryear{Neese}{Neese}{2022}]{Neese2022}
Neese F.,  2022, \mn@doi [WIREs Computational Molecular Science]
  {10.1002/wcms.1606}, 12

\bibitem[\protect\citeauthoryear{Neese, Wennmohs, Becker  \& Riplinger}{Neese
  et~al.}{2020}]{Neese2020}
Neese F.,  Wennmohs F.,  Becker U.,   Riplinger C.,  2020, \mn@doi [Journal of
  Chemical Physics] {10.1063/5.0004608}, 152, 224108

\bibitem[\protect\citeauthoryear{Noble, Dulieu, Congiu  \& Fraser}{Noble
  et~al.}{2011}]{noble_co_2011}
Noble J.~A.,  Dulieu F.,  Congiu E.,   Fraser H.~J.,  2011, \mn@doi [The
  Astrophysical Journal] {10.1088/0004-637X/735/2/121}, 735, 121

\bibitem[\protect\citeauthoryear{Oba, Watanabe, Kouchi, Hama  \&
  Pirronello}{Oba et~al.}{2010a}]{oba_experimental_2010}
Oba Y.,  Watanabe N.,  Kouchi A.,  Hama T.,   Pirronello V.,  2010a, \mn@doi
  [The Astrophysical Journal] {10.1088/2041-8205/712/2/L174}, 712, L174

\bibitem[\protect\citeauthoryear{Oba, Watanabe, Kouchi, Hama  \&
  Pirronello}{Oba et~al.}{2010b}]{Oba2010Carbonic}
Oba Y.,  Watanabe N.,  Kouchi A.,  Hama T.,   Pirronello V.,  2010b, \mn@doi
  [The Astrophysical Journal] {10.1088/0004-637X/722/2/1598}, 722, 1598

\bibitem[\protect\citeauthoryear{Papajak, Zheng, Xu, Leverentz  \&
  Truhlar}{Papajak et~al.}{2011}]{papajak_perspectives_2011}
Papajak E.,  Zheng J.,  Xu X.,  Leverentz H.~R.,   Truhlar D.~G.,  2011,
  \mn@doi [J. Chem. Theory. Comp.] {10.1021/ct200106a}, 7, 3027

\bibitem[\protect\citeauthoryear{Pauly \& Garrod}{Pauly \&
  Garrod}{2018}]{Pauly2018}
Pauly T.,  Garrod R.~T.,  2018, \mn@doi [The Astrophysical Journal]
  {10.3847/1538-4357/aaa96a}, 854, 13

\bibitem[\protect\citeauthoryear{Perdew, Ruzsinszky, Csonka, Constantin  \&
  Sun}{Perdew et~al.}{2009}]{perdew_workhorse_2009}
Perdew J.~P.,  Ruzsinszky A.,  Csonka G.~I.,  Constantin L.~A.,   Sun J.,
  2009, \mn@doi [Physical Review Letters] {10.1103/PhysRevLett.103.026403},
  103, 026403

\bibitem[\protect\citeauthoryear{Perrero, {Enrique-Romero},
  {Mart{\'i}nez-Bachs}, Ceccarelli, Balucani, Ugliengo  \& Rimola}{Perrero
  et~al.}{2022}]{Perrero2022}
Perrero J.,  {Enrique-Romero} J.,  {Mart{\'i}nez-Bachs} B.,  Ceccarelli C.,
  Balucani N.,  Ugliengo P.,   Rimola A.,  2022, \mn@doi [ACS Earth and Space
  Chemistry] {10.1021/acsearthspacechem.1c00369}, 6, 496

\bibitem[\protect\citeauthoryear{Qasim, Lamberts, He, Chuang, Fedoseev,
  Ioppolo, Boogert  \& Linnartz}{Qasim et~al.}{2019}]{Qasim2019}
Qasim D.,  Lamberts T.,  He J.,  Chuang K.~J.,  Fedoseev G.,  Ioppolo S.,
  Boogert A.~C.,   Linnartz H.,  2019, \mn@doi [Astronomy and Astrophysics]
  {10.1051/0004-6361/201935068}, 626, A118

\bibitem[\protect\citeauthoryear{Rappoport \& Furche}{Rappoport \&
  Furche}{2010}]{rappoport_property-optimized_2010}
Rappoport D.,  Furche F.,  2010, \mn@doi [J. Chem. Phys.] {10.1063/1.3484283},
  133, 134105

\bibitem[\protect\citeauthoryear{Rimola, Taquet, Ugliengo, Balucani  \&
  Ceccarelli}{Rimola et~al.}{2014}]{Rimola2014}
Rimola A.,  Taquet V.,  Ugliengo P.,  Balucani N.,   Ceccarelli C.,  2014,
  \mn@doi [Astronomy and Astrophysics] {10.1051/0004-6361/201424046}, 572

\bibitem[\protect\citeauthoryear{Rocha et~al.,}{Rocha
  et~al.}{2024}]{rocha_jwst_2024}
Rocha W. R.~M.,  et~al., 2024, \mn@doi [Astronomy \& Astrophysics]
  {10.1051/0004-6361/202348427}, 683, A124

\bibitem[\protect\citeauthoryear{Sanz-Novo et~al.,}{Sanz-Novo
  et~al.}{2023}]{sanz-novo_discovery_2023}
Sanz-Novo M.,  et~al., 2023, \mn@doi [The Astrophysical Journal]
  {10.3847/1538-4357/ace523}, 954, 3

\bibitem[\protect\citeauthoryear{Senevirathne, Andersson, Dulieu  \&
  Nyman}{Senevirathne et~al.}{2017}]{SENEVIRATHNE201759}
Senevirathne B.,  Andersson S.,  Dulieu F.,   Nyman G.,  2017, \mn@doi
  [Molecular Astrophysics] {10.1016/j.molap.2017.01.005}, 6, 59

\bibitem[\protect\citeauthoryear{Senosiain, Klippenstein  \& Miller}{Senosiain
  et~al.}{2005}]{senosiain_complete_2005}
Senosiain J.~P.,  Klippenstein S.~J.,   Miller J.~A.,  2005, \mn@doi
  [Proceedings of the Combustion Institute] {10.1016/j.proci.2004.07.009}, 30,
  945

\bibitem[\protect\citeauthoryear{Sure \& Grimme}{Sure \&
  Grimme}{2013}]{Sure2013}
Sure R.,  Grimme S.,  2013, \mn@doi [Journal of Computational Chemistry]
  {10.1002/jcc.23317}, 34, 1672

\bibitem[\protect\citeauthoryear{Suzuki, Furuya, Aikawa, Shibata  \&
  Majumdar}{Suzuki et~al.}{2024}]{suzuki_chemical_2024}
Suzuki T.,  Furuya K.,  Aikawa Y.,  Shibata T.,   Majumdar L.,  2024, \mn@doi
  [Monthly Notices of the Royal Astronomical Society] {10.1093/mnras/stae1589},
  532, 1796

\bibitem[\protect\citeauthoryear{Terwisscha~van Scheltinga, Ligterink, Bosman,
  Hogerheijde  \& Linnartz}{Terwisscha~van Scheltinga
  et~al.}{2022}]{terwisscha_van_scheltinga_formation_2022}
Terwisscha~van Scheltinga J.,  Ligterink N. F.~W.,  Bosman A.~D.,  Hogerheijde
  M.~R.,   Linnartz H.,  2022, \mn@doi [Astronomy \& Astrophysics]
  {10.1051/0004-6361/202142181}, 666, A35

\bibitem[\protect\citeauthoryear{Wakelam et~al.,}{Wakelam
  et~al.}{2017}]{Wakelam2017h2}
Wakelam V.,  et~al., 2017, \mn@doi [Molecular Astrophysics]
  {10.1016/j.molap.2017.11.001}, 9, 1

\bibitem[\protect\citeauthoryear{Weigend \& Ahlrichs}{Weigend \&
  Ahlrichs}{2005}]{Weigend2005}
Weigend F.,  Ahlrichs R.,  2005, \mn@doi [Physical Chemistry Chemical Physics]
  {10.1039/b508541a}, 7, 3297

\bibitem[\protect\citeauthoryear{Zhao \& Truhlar}{Zhao \&
  Truhlar}{2005}]{Zhao2005}
Zhao Y.,  Truhlar D.~G.,  2005, \mn@doi [J. Phys. Chem. A] {10.1021/jp050536c},
  109, 5656

\makeatother
\end{thebibliography}




\appendix

\section{Exploring the bounds of R.6} \label{sec:app1}

In the main text, for reaction \Reacref{eq:c_co} (\ce{c-HOCO + H -> CO + H2O}) on the ASW ice cluster we calculate an upper bound of the activation energy owing to the impossibility of converging a first order saddle point as the structural TS. On the contrary, we find that a torsional constrain is necessary to converge the structural TS to a second order saddle point with the second imaginary frequency corresponding to the torsional angle leading to \ce{c-HOCO -> t-HOCO} isomerization. Therefore, and because reaction \Reacref{eq:t_co} (the equivalent reaction to \Reacref{eq:c_co} in the t-HOCO conformer) is barrierless it is fair to consider whether providing an ``upper bound'' for the barrier of reaction \Reacref{eq:c_co} or if rather \Reacref{eq:c_co} is barrierless due to spontaneous torsion. 

\begin{figure}
	\centering
	\includegraphics[width=0.5\linewidth]{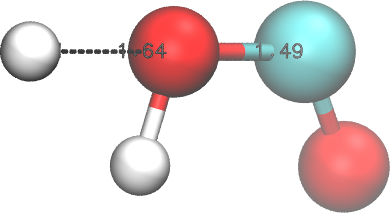}
	\caption{Transition state for the gas-phase, i.e. in the absence of a water matrix for the \ce{c-HOCO + H -> H2O + CO} reaction. The H-O-H angle is 78.6$^{\circ}$}
	\label{fig:app1}
\end{figure}

While certainly not being able to conclusively determine a proper first order saddle point is an obvious caveat of our simulations, we have arguments to believe that such a first order saddle point should exist and that the presence of a small second imaginary frequency is an artifact of the optimizer or the potential, i.e. not a real physical effect. Our reason to believe such an outcome stems from the fact that we \emph{can} isolate the first order saddle point in the gas-phase, that is, without additional water molecules. This transition state, calculated using the benchmark level revTPSSh(D3BJ)/def2-TZVPPD is shown in Figure \ref{fig:app1}. The $\Delta U_{A}$ calculated for this structure and a reactant state at a large HOCO-H distance reveals a value of the height of the barrier, ZPVE inclusive, of $\Delta U_{A}$=4.6 \kcalmol.\footnote{In Section \ref{sec:method:bs} and Figure \ref{fig:benchmark} we report 5.3 \kcalmol at the CASPT2(18,14)/cc-pVTZ//MPWB1K(D3BJ)/def2-TZVPPD level which is very similar to the values presented in the appendix with the best performing DFT method (See Section \ref{sec:method:bs}).} This value is certainly lower than the upper bounds on the 18 \ce{H2O} cluster, which is coherent. Besides, this value is also lower than the gas-phase isomerization barrier calculated in the main text (6.4 \kcalmol; Section \ref{sec:isomers}). Therefore, we find it difficult to conceive that, with a torsional barrier higher than the hydrogenation one, the water matrix can exert an influence so high to make torsion spontaneous. Nonetheless, we encourage further investigation of this issue by other groups.


\bsp	
\label{lastpage}
\end{document}